\documentclass[aps,pre,onecolumn,showpacs,showkeys]{revtex4}
\input epsf  
\usepackage{graphicx} 
\usepackage{amsmath}
\usepackage{amssymb}
\usepackage{amscd}

\newcommand{\Li}{\mathrm{Li}}

\begin{document} 

 \title{The Asymmetric Simple Exclusion Process~:
 An Integrable Model for  Non-Equilibrium Statistical Mechanics}
\author{Olivier Golinelli$^{1}$ and Kirone Mallick$^{1,2}$}
  \affiliation{ (1) Service de Physique Th\'eorique, C. E. A.  Saclay,
 91191 Gif-sur-Yvette Cedex, France \\
                (2) Isaac Newton Institute for Mathematical Sciences,
  20 Clarkson Road, Cambridge  CB3 0EH, UK }
  \date{23 June 2006, in final form 9 August 2006, published 27
        September 2006}

\begin{abstract} 
   The Asymmetric Simple Exclusion Process
  (ASEP)   plays the role of a paradigm in  
 Non-Equilibrium Statistical Mechanics. We  review exact results
 for  the ASEP obtained by  Bethe Ansatz and put 
 emphasis   on the algebraic properties of this model.
 The  Bethe equations for  the eigenvalues  of the Markov Matrix of the ASEP 
 are derived  from 
 the algebraic Bethe Ansatz. Using  these  equations
 we explain how to  calculate  the  spectral gap of the model
 and how global spectral properties such as the existence of multiplets
 can be predicted. An extension of the Bethe Ansatz leads to an analytic
 expression for the large deviation function of the current in the ASEP 
 that satisfies the Gallavotti-Cohen relation. Finally, we describe
 some  variants of the ASEP  that are also  solvable by  Bethe Ansatz. 
  \end{abstract}
\maketitle

 \section{Introduction}

 Equilibrium statistical mechanics tells us that the  
 probability  of a microstate  of  
  a  system  in equilibrium with a thermal reservoir 
  is given by  the Boltzmann-Gibbs   
  law~: if  ${\mathcal H}$ is  the Hamiltonian  of the system,
  the probability distribution over the configuration
 space is proportional to $\exp(-\beta  {\mathcal H})$  where
 $\beta$ is the inverse of temperature. This 
  {\it canonical}  prescription  is the  starting point for any  study
 of a  system in thermodynamic equilibrium~: it has provided a firm
  microscopic basis for   the 
  laws of classical thermodynamics,   has allowed us 
 to describe various states of matter (from liquid crystals
 to superfluids), and has led to a deep understanding of phase
 transitions that culminated in  the renormalisation group theory.  

 For a  system out of equilibrium, the probability of 
 a given  microstate  evolves constantly with time. In the long time limit 
 such a   system may reach a stationary state 
  in which the  probability measure  over the configuration
 space converges to a well-defined  and constant distribution. If the system
  carries macroscopic stationary  currents
 that represent  interactions and 
 exchanges of matter  or energy  with the external
 world, this stationary   distribution  is  {\it generically not}
  given   by  the canonical  Boltzmann-Gibbs  law.  At present,
 there exists no theory that can predict the 
  stationary state of a system far from equilibrium  from a knowledge of 
 the microscopic interactions of the elementary constituents
 of the system  amongst themselves and with  the external
 environment,  and the dynamical rules  that govern its evolution.  The 
 search for general features and laws of {\it non-equilibrium 
 statistical mechanics} is indeed  a central topic of  current research. 

  For systems close  to  thermodynamic equilibrium,    linear-response  theory 
  yields  the fluctuation-dissipation relations
 and the Onsager reciprocity relations. However, these properties 
  of non-equilibrium  statistical mechanics are only
   valid  in the vicinity of  equilibrium. 
   Another  route to explore and discover the characteristics
  of a general theory for systems out of equilibrium is through
 the  study of   mathematical models. These models
 should  be simple enough to allow   an exact and thorough
 mathematical analysis, but at the same time they must exhibit
 a complex phenomenology and must be versatile enough to be
 of physical significance. 
 In  the  theory of phase transitions such a corner-stone
 was provided  by the Ising model. It is  natural to expect that a 
  well-crafted  dynamical version    of  
   the Ising model could   play a key role in the development
 of  non-equilibrium  statistical mechanics.  Driven lattice gases
 (Katz, Lebowitz and Spohn 1984; for a review see
 Schmittmann and  Zia, 1995) provide such   models; they represent 
  particles hopping  on a lattice  and  interacting   through
 hard-core exclusion. The particles 
  are  subject to  an external field that induces  a bias
 in the hopping rates  resulting in   a stationary  macroscopic
 current in  the system. Due to  this current, the microscopic
 detailed balance condition is violated and the system is 
 generically in a non-equilibrium stationary state.

    The Asymmetric Simple  Exclusion Process (ASEP) 
 in one dimension  is a driven lattice gas that  can be viewed
 as a special case of the general class of models defined 
 by Katz, Lebowitz and Spohn. This stochastic  particle system
 was simultaneously introduced as  a biophysical  model for
 protein synthesis on RNA   (MacDonald and Gibbs 1969) 
 and as a purely mathematical tool  for the study of interaction
 of Markov processes (Spitzer 1970, Liggett 1985, Spohn 1991, Liggett 1999). 
 Subsequently, the ASEP  has been used to study a wide range
 of physical phenomena~:
 hopping conductivity in  solid electrolytes (Richards 1977),
 transport of Macromolecules through thin vessels 
 (Levitt 1973),  reptation of polymer in a gel (Widom et al. 1991), 
   traffic  flow  (Schreckenberg and Wolf 1998), surface growth 
 (Halpin-Healy and Zhang 1995, Krug 1997), 
  sequence alignment (Bundschuh 2002) and molecular motors (Klumpp
 and Lipowsky, 2003).

    From a theoretical   point of view,
the ASEP plays  a fundamental role  in the study of non-equilibrium
processes (Krug 1991).
  Many exact results for the ASEP  have been derived  using two
  complementary approaches, the Matrix Product Ansatz and the Bethe Ansatz 
  (for a review see Derrida 1998, Sch\"utz 2001),  the   relation
between these two  Ans\"atze   being  still a matter of investigation 
 (Stinchcombe and Sch\"utz 1995; Alcaraz and Lazo 2004).
 
  The Matrix Product Ansatz, inspired from the quantum inverse
 scattering method (Faddeev  1984),    is based on 
  a representation  of the components of the steady  state wave function
 of the Markov operator  in terms of a product of non-commuting
  operators. This Matrix Product technique, initially
   introduced for the ASEP with open boundaries  (Derrida et al. 1993), 
   has proved to be  a very efficient tool  to  calculate steady state
  properties  such as equal time correlations in the steady state,
   current fluctuations   (Derrida et al. 1997),   and large 
 deviation functionals  (Derrida et al. 2003). This technique  
  has been used  to prove  rigorously  that 
  the invariant distribution of the  ASEP with  two classes  of particles
  is not a Gibbs measure (Speer 1993).  

 The second approach, which consists in 
 applying the Bethe Ansatz to a non-equilibrium
 process such as the ASEP,  is due  to  D. Dhar (1987). The Markov
 matrix that encodes the stochastic dynamics of the ASEP
 can be rewritten  in terms  of Pauli matrices;   in 
 the absence of a driving field,
the {\it symmetric} exclusion process can be mapped exactly  into the
Heisenberg spin chain.  The asymmetry due to a non-zero external
driving field breaks the left/right symmetry and the ASEP becomes 
equivalent to a non-Hermitian spin chain of the XXZ type
 with boundary terms that preserve the integrable 
 character of the model.  The ASEP can also be mapped
 into a six vertex model (Baxter 1982, Kandel et al., 1990).
 These mappings  suggest the  use of  the Bethe Ansatz
 to derive spectral information about 
  the evolution operator, such as the spectral
 gap (Gwa and Spohn 1992, Kim 1995, Golinelli and Mallick 2004a, 2005a)
 and large deviation functions 
(Derrida and Lebowitz
1998; Derrida and Appert 1999; Derrida and Evans 1999).

 The aim of the present work is to provide  the reader with 
 an introduction to integrability methods  applied to the
 ASEP and to describe some of the exact results derived with
 the help of the Bethe  Ansatz. Our  presentation 
 assumes little prior knowledge of these methods and we  put 
 emphasis on the algebraic properties of the model.  In fact, 
  the ASEP on a periodic ring is  one of the most elementary systems  to which
 the   Bethe  Ansatz can be  applied~: it is a discrete and classical
 model of interacting particles with the simplest possible
 interaction, hard-core exclusion, and with a  conservation law
 (the total number of particles is constant).
 
  The layout of this work is as follows~:
 in section~\ref{sec:SYM},    we describe
  the natural  symmetries  (by translation,
 reflection and charge-conjugaison) of the ASEP on a periodic ring; 
  in section~\ref{sec:ABA}, we 
 review the solution of  the ASEP by   algebraic
 Bethe Ansatz and derive the Bethe equations that determine the spectrum
 of the model. In the case of the totally  asymmetric simple 
 exclusion process, the analysis of the  Bethe equations can be
 carried out very precisely, even for finite size systems~:
 in section~\ref{sec:eqTASEP}, we explain the procedure
 for  solving the  TASEP Bethe equations,  determine
 the spectral gap, and show that a hidden symmetry of these equations
 allows  to predict the existence of  unexpected spectral degeneracies  
 in the spectrum. In section~\ref{sec:LDF}, we  discuss the method for 
  calculating   the large deviation function of the total current;
 in particular, we show that the Gallavotti-Cohen relation manifests itself
 as a symmetry of the Bethe equations. In the last section~\ref{sec:variantes},
 we review the applications of  integrability techniques to some  variants
 of the ASEP. 

 \section{Basic Properties  of the Exclusion Process}
\label{sec:SYM}

 \subsection{Definition of the model~: The Markov matrix}

\begin{figure}[th]
  \includegraphics[height=5.0cm]{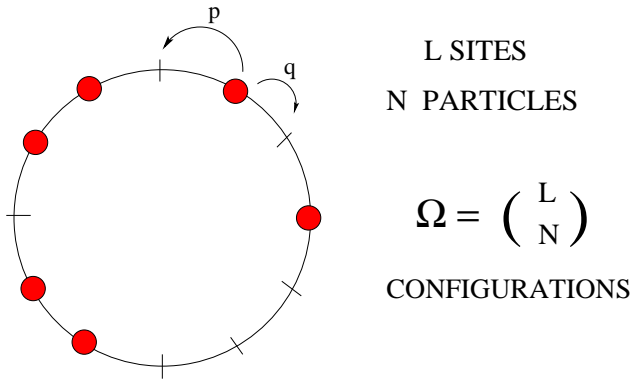}
   \caption{The Asymmetric exclusion process on a periodic
 ring. }
  \label{fig:periodicASEP}
\end{figure}

   We consider the exclusion process   on a periodic
    one dimensional lattice with $L$ sites (sites $i$ and $L + i$ are
   identical).  A  lattice site cannot be  occupied by more than  one particle.
   The state of  a site  $i$ ($1 \le i \le L$)  is characterized
  by  a  Boolean number $\tau_i = 0$ or $ 1$ according as  the site  $i$ is
   empty or occupied. A configuration $\mathcal{C}$  can be  represented  by 
 the sequence $(\tau_1, \tau_2, \ldots, \tau_L).$
  The system evolves   with  time according to 
the following stochastic rule (see Figure~\ref{fig:periodicASEP}):
 a particle on a site $i$ at time $t$ jumps, in
the interval between  $t$ and $t+dt$, with probability $p\ dt$ to the
neighbouring site $i+1$ if this site is empty ({\em exclusion rule})
 and with  probability $q\ dt$ to the
 site $i-1$ if this site is empty. The jump  rates $p$ and $q$
 are normalized such that  $ p + q =1$. In 
  the totally asymmetric exclusion process (TASEP) 
 the jumps are totally biased in one direction ($p =1$ or  $q =1$).

  We call $P_t(\mathcal{C})$
 the probability of configuration $\mathcal{C}$ at time $t$.
  As the exclusion process is a continuous-time Markov process, the time
 evolution of $P_t(\mathcal{C})$ is determined by the master equation 
\begin{equation}
    \frac{d}{dt} P_t(\mathcal{C})  = \sum_{\mathcal{C}'}
      M(\mathcal{C},\mathcal{C}') P_t(\mathcal{C}') 
   =    \sum_{\mathcal{C}'}   
 \Big(  M_0(\mathcal{C},\mathcal{C}') +   M_1(\mathcal{C},\mathcal{C}')
+   M_{-1} (\mathcal{C},\mathcal{C}')\Big)
          P_t(\mathcal{C}')   \, . 
 \label{eq:Markov}
\end{equation}
 The   Markov  matrix  $M$  encodes  the dynamics of the exclusion process:  
 the  non-diagonal  element  $ M_1(\mathcal{C},\mathcal{C}')$
 represents
 the  transition rate  from configuration $\mathcal{C}'$ to $\mathcal{C}$
 where  a particle hops  in the  forward ({\it i.e.}, anti-clockwise)
   direction,
  the  non-diagonal  element  $ M_{-1}(\mathcal{C},\mathcal{C}')$
 represents
 the  transition rate  from configuration $\mathcal{C}'$ to $\mathcal{C}$
where  a particle hops  in the  backward  ({\it i.e.}, clockwise)  direction.
 The diagonal term $M_0(\mathcal{C},\mathcal{C}) =
  -  \sum_{\mathcal{C}'\neq \mathcal{C}  }\left(  M_1(\mathcal{C}',\mathcal{C})
  +   M_{-1} (\mathcal{C}',\mathcal{C}) \right)   $ 
represents  the exit rate
 from  the configuration $\mathcal{C}$.

The   matrix $M$ is a real non-symmetric matrix and, therefore, its
eigenvalues (and eigenvectors) are either real numbers or complex
conjugate pairs.
 A right eigenvector $P$ is associated with the eigenvalue $E$
of $M$ if
\begin{equation}
  M P = EP      \, . 
  \label{eq:mpsi=epsi}
\end{equation}
 On a ring the total number $N$ of particles  is conserved. For a given
 value of $N$, 
 the dynamics is ergodic ({\it i.e.}, $M$ is 
irreducible and aperiodic). The Perron-Frobenius theorem
implies that 
 0 is a non-degenerate eigenvalue and that all other 
eigenvalues   have a strictly negative real
part; the relaxation time of the corresponding eigenmode is $\tau =
-1/\mathrm{Re}(E)$.  The    right eigenvector
 corresponding   to the eigenvalue $E = 0$
 is the stationary state~:
 for the ASEP
 on a ring the steady state    is uniform and 
 the stationary probability of  any configuration 
  $\mathcal{C}$   is given by 
 $P(\mathcal{C}) = N!(L-N)!/L!$.

 {\it Remark~:} A configuration can  also be characterized
  by the positions of the $N$ particles
on the ring, $(x_1, x_2, \dots, x_N)$ with $1 \le x_1 < x_2 < \dots < x_N
\le L$. With  this representation, the eigenvalue equation
 (\ref{eq:mpsi=epsi})   becomes
\begin{eqnarray}
 &&  E P(x_1,\dots, x_N) =   \nonumber \\   &&
  \sum_i  p \left[ 
             P(x_1, \dots, x_{i-1},\ x_i-1,\ x_{i+1}, \dots, x_N) 
 - P(x_1,\dots, x_n) \right] + 
    \nonumber \\      
  &&  \sum_j  q \left[ P(x_1, \dots, x_{j-1},\ x_j+1,\ x_{j+1}, \dots, x_N)
              - P(x_1,\dots, x_N) \right] \, , 
\end{eqnarray}
where the sum runs over the indexes $i$ such that $ x_{i-1} < x_i-1$
 and   over the indexes $j$ such that $  x_j + 1 < x_{j+1} \,;$   these
 inequalities ensure  that the corresponding jumps  are  allowed.

   The knowledge of the spectrum of the matrix $M$ and of the
 associated eigenvectors  provides 
 a full description of the dynamics of the model. Such
 a spectral analysis is very similar to that  carried out
 in Quantum Mechanics. One should however keep in mind that
 in the present case  the evolution operator is non-hermitian~:
 its  eigenvalues are  complex numbers and the associated
 eigenvectors do not form  an orthogonal basis in
 the configuration space. Moreover, because $M$ is not
 symmetric, its left eigenvectors are different from
 the right eigenvectors.

 \subsection{Natural symmetries of ASEP on a periodic ring}
 \label{sec:sym}

 Before   diagonalizing the  Markov  matrix $M$ by Bethe Ansatz,
 we present  some  invariance properties of the ASEP, that can be described
 using elementary methods.   The exclusion
 process on a ring displays  symmetry properties
 under translation, right-left reflection, and particle/hole
 exchange (the last two  symmetries play a role  analogous to parity
 and charge conjugaison in quantum mechanics).  These    symmetries 
  reveal some intrinsic  formal properties of the  Markov  matrix, and 
 are  very useful  to reduce the  computation time when  
  numerical diagonalization is performed.  Finally,  these
 natural symmetries will allow us to define some conserved
 quantities (analogous to  quantum numbers) and   to predict
 degeneracies in  the spectrum.

\subsubsection{Translation invariance}
 \label{sec:Tinv}

 The translation operator $T$ shifts  simultaneously  all the particles 
  one site forward: 
\begin{equation}
   T (\tau_1, \tau_2, \ldots, \tau_L)  = 
 (\tau_L, \tau_1, \tau_2, \ldots, \tau_{L-1})    \, .
 \label{eq:defT}
\end{equation}
Because of  the periodicity  we have $T^L = 1$.
  Thus,  the eigenvalues (or impulsions)  $k$ of $T$
 are simply  the $L$-th roots of unity:
 \begin{equation}
  k  =  e^{2i\pi m/L} \ \ \, \hbox{ for } m = 0, \ldots, L -1 \, .
\label{eq:eigenvalT}
\end{equation}
 We denote by  ${\bf T}(k)$ the 
 eigenspace of $T$  corresponding  to  the impulsion $k$.
 The projection  operator $\mathcal{T}_k$  over  this eigenspace is given by 
 \begin{equation}
  \mathcal{T}_k = \frac{1}{L}  \sum_{r=0}^{L-1}  k^{-r} T^r  \, .
\label{eq:projector}
\end{equation}
 The complex conjugation transforms a $T$-eigenvector  of 
 eigenvalue $k$ into an eigenvector  with
 eigenvalue $k^* = k^{-1}$, {\it i.e.},  ${\bf T}(k^*) = [{\bf T}(k)]^*$.

 The ASEP on a periodic lattice is translation invariant, {\it i.e.},
 \begin{equation}
  M  T  = T M  \, .
 \label{eq:MT=TM}
\end{equation}
The  matrix $M$  and the translation operator $T$ can therefore
 be simultaneously diagonalized:  $M$  
 leaves  each  eigenspace ${\bf T}(k)$  invariant.  We denote
 by ${\bf Sp}(k)$  the 
 set of the eigenvalues of  $M$ restricted to ${\bf T}(k)$.
 Using  complex conjugation we obtain 
 the property
 \begin{equation}
       {\bf Sp}(k^*) =   \left[{\bf Sp}(k) \right]^*   \, .
\label{eq:Spconj}
\end{equation}

\subsubsection{Right-Left Reflection}

  The  reflection operator $R$  interchanges    the right and the left 
 and  is defined by 
\begin{equation}
   R (\tau_1, \tau_2, \ldots, \tau_L) 
  =  (\tau_L, \tau_{L-1}, \ldots, \tau_2, \tau_1)           \, .
 \label{eq:defR}
\end{equation}
 We have $R^2 = 1$;  the eigenvalues of $R$
 are thus $r = \pm 1$. The reflection  $R$ reverses
   the translations,  {\it i.e.},
  \begin{equation}
 RT = T^{-1} R  \, ,
 \label{eq:RT}
\end{equation}
 and transforms a $T$-eigenvector  of 
 eigenvalue $k$ into an eigenvector  of 
 eigenvalue $k^*$.
  The operator  $R$ does not commute with the Markov matrix  because 
  the  asymmetric jump  rates  $(p \neq q)$ are  not invariant 
 under  the  exchange  of  right and left.
 Writing explicitly
 the dependence of $M$ on the jump rates, we have 
\begin{equation}
  R M_{(p, q)} R^{-1} =  M_{(q, p)}          \, .
 \label{eq:RMR}
\end{equation}
 Thus,   in general the reflection is {\it not} a symmetry of the ASEP
  (only  the symmetric exclusion process is invariant under  reflection).

\subsubsection{Charge conjugation}

  The charge conjugation operator $C$ exchanges particles and holes
 in the system, {\it i.e.}, a configuration with $N$ particles
 is mapped into  a configuration with $L - N$ particles: 
\begin{equation}
   C (\tau_1, \tau_2, \ldots, \tau_L) 
  =  ( 1- \tau_1, 1 - \tau_2, \ldots, 1 - \tau_L)             \, .
 \label{eq:defC}
\end{equation}
 The operator $C$ satisfies the relations: 
\begin{equation}
 C^2 = 1 , \, \,\,\,\,  CT = TC, \,\,\,\,\, CR = RC \, .
\label{eq:C2}
\end{equation}
 By charge conjugation,  particles 
  jumping forward 
  are  mapped  into holes  jumping forward. But holes
  jumping forward are equivalent to 
   particles jumping backward.  Writing explicitly
 the dependence of $M$  on  the number $N$  of particles
 and  on the jump rates, we thus have 
\begin{equation}
  C M_{(N, p, q)} C^{-1} =  M_{(L-N, q, p)}          \, .
 \label{eq:CMC}
\end{equation}
 We notice  that the number of particles is conserved by $C$ only at half
 filling ($L = 2N $). 
  Thus,   the charge conjugation is {\it not} a symmetry of the ASEP
  except  for the symmetric $(p=q)$ exclusion process at half filling.

\subsubsection{CR symmetry}
\label{sec:CR}

  Using   equations~(\ref{eq:RMR}) and~(\ref{eq:CMC})
 to  combine  the charge
conjugation $C$ with the reflection $R$, we obtain 
\begin{equation}
  (CR) M_{(N, p, q)} (CR)^{-1} =  M_{(L-N, p, q)}         \, .
 \label{eq:CRM}
\end{equation}
 Thus,  for a given $L$, the  $CR$ operator
  maps  the ASEP with $N$ particles into the
 ASEP with  the same jumping rates  but with $L-N$ particles.  This implies 
  that the spectrum of $M$ for $N$ particles is identical with  the spectrum
for $L-N$ particles (because  $CR$ transforms eigenvectors of $M_{(N,p,q)}$
 into eigenvectors of  $M_{(L-N,p,q)}$).

 For  the ASEP model
 at half filling, {\it i.e.}, $L
 = 2N$,   the  $CR$ operator   
 constitutes an exact symmetry~:
 \begin{equation}
  (CR) \, M = M \, (CR)  \, .
 \end{equation}
 The ASEP at half filling is therefore invariant 
under each of the two symmetries, 
translation $T$ and $CR$.  Note  that they do not commute with each
other;  we  rather obtain from equations~(\ref{eq:RT}) and~(\ref{eq:C2})
\begin{equation}
  (CR) T = T^{-1} (CR) \, . 
 \label{eq:CRT}
\end{equation}
Hence,  the $CR$   transformation  maps the subspace ${\bf T}(k)$
into ${\bf T}(k^*)$ and thus
     $  {\bf Sp}(k) =   {\bf Sp}(k^*)   \, .$
For $k \neq \pm 1$, the  subspaces ${\bf T}(k)$ and ${\bf T}(k^*)$ are
distinct and  the corresponding
eigenvalues of $M$ are doubly degenerate.  For $k = \pm 1$, the $CR$
transformation leaves ${\bf T}(k)$ invariant. Thus, the natural symmetries
do not predict any degeneracies in the sets ${\bf Sp}(\pm 1)$. However,
each of the subspaces ${\bf T}(\pm 1)$ is split  into two smaller
subspaces which are  invariant under $CR$ and on which $CR = \pm 1$.
  For the   ASEP at half filling, we find   from  Equation~(\ref{eq:Spconj})
  that   the set ${\bf Sp}(k)$ is 
self-conjugate, for all $k$,  and degenerate with  ${\bf Sp}(k^*)$ 
\begin{equation} 
 {\bf Sp}(k) = \left[  {\bf Sp}(k) \right]^* =   {\bf Sp}(k^*) \, .
\end{equation}
This means that ${\bf Sp}(k)$, for all $k$, is made only of 
real numbers or  of complex conjugate pairs.

 \section{Algebraic Bethe Ansatz for the Exclusion Process}
\label{sec:ABA}

  \subsection{The algebra of local update operators}

  The space of all  possible configurations
 ${C}$ is a     $2^L$ dimensional vector space
 that we shall denote by   $\mathcal{S}$.
 Each site  $i$  is  either occupied or empty~:
 we shall  represent its  state  by the local basis
   $(|1\rangle_i, |0\rangle_i)$   of the two dimensional space
   ${\mathbb  C}^2_i$. The total configuration  space  $\mathcal{S}$
 is thus given by 
 \begin{equation}
  \mathcal{S} =  {\mathbb  C}^2_1 \otimes {\mathbb  C}^2_2 \otimes 
  \ldots    \otimes   {\mathbb  C}^2_L \, , 
\label{eq:defS}
\end{equation}
 The natural basis of   the  configuration space    $\mathcal{S}$
 is the tensor product  $(|1\rangle_1, |0\rangle)_1 \otimes  \ldots  \otimes
  (|1\rangle_L, |0\rangle)_L$.
On a periodic lattice, 
 the  number $N$ of particles is conserved by the dynamics. The total number of
configurations for $N$ particles on a ring with $L$ sites is given by
$\mathcal{ N } = L! / [ N! (L-N)!]$. The  allowed configurations of the
 system  span a subspace of  $\mathcal{S}$ of dimension 
 $\mathcal{ N }$.  The Markov matrix can be expressed
 as a sum of local  operators that  update  the
 bond $(i, i+1)$~:
\begin{equation}
             M =  \sum_{i=1}^{L} M_{i, i+1}  \, , .  
\label{eq:sumlocalM}
\end{equation}
 The    update operator  $ M_{i, i+1}$ is given by
\begin{equation}
         M_{i, i+1}  = 
 {\bf 1}_1 \otimes {\bf 1}_2 \ldots {\bf 1}_{i-1}  \otimes
  \left( \begin{array}{cccc}
                         0 &  0  &  0 &  0  \\
                         0 & - p &  q &  0  \\
                         0 &   p & -q &  0  \\
                         0 &  0  &  0 &  0  
                   \end{array}
                   \hspace{0.2in} \right)  
         \otimes {\bf 1}_{i+2} \ldots {\bf 1}_L \, ,
\label{eq:localM}
\end{equation}
 where  ${\bf 1}_{j}$ is the $2 \times 2$ identity matrix
 acting on the site number $j$.
 We emphasize that  $ M_{i, i+1}$ is   a  $2^L\times 2^L$ matrix
 that acts trivially on all sites different from $i$ and $i+1$
 and updates  the bond  $(i, i+1)$ according to the local
 dynamical rules of the exclusion process. 
 We also define the local permutation operator   $P_{i,i+1}$
\begin{equation}
         P_{i, i+1}  = 
 {\bf 1}_1 \otimes {\bf 1}_2 \ldots {\bf 1}_{i-1}  \otimes
  \left( \begin{array}{cccc}
                         1 &  0  &  0 &  0  \\
                         0 &  0 &  1 &  0  \\
                         0 &  1 &  0 &  0  \\
                         0 &  0  &  0 &  1  
                   \end{array}
                   \hspace{0.2in} \right)  
         \otimes {\bf 1}_{i+2} \ldots {\bf 1}_L \, .
\label{eq:localP}
\end{equation}

 The local update operators satisfy the following relations:
\begin{eqnarray}
     P_{i,i+1} M_{i, i+1}  &=&  -  M_{i, i+1}   \,  \label{eq:PM}  \\
     M_{i, i+1}^2  &=&       -  M_{i, i+1}  \, \label{eq:M2} \\
     M_{i, i+1}\,  M_{i+1, i+2} \,  M_{i, i+1} &=& pq \, M_{i, i+1}  \, 
      \label{eq:MMM} \\
       M_{i, i+1}\,  M_{i-1, i} \,  M_{i, i+1} &=& pq \, M_{i, i+1}  \, 
      \label{eq:MMMb} \\ 
   \,  [M_{i,i+1}, M_{j,j+1}]    &=&  0   \,\,\, \hbox{ for } |i -j| \ge 2 \, .
   \label{eq:MiMj}
\end{eqnarray}
 These identities define a Temperley-Lieb algebra~: this 
  property of   reaction-diffusion processes 
 was  emphasized by Alcaraz et al. (1994) and  plays a key-role
 in the integrability of the ASEP.

 \subsection{The ASEP as  a non-Hermitian spin chain} 
 
 The dynamics of the ASEP is entirely encoded in its Markov matrix
 $M$ that governs the time evolution of the probability 
 measure on the configuration space. We  now show that the local 
  operator  $M_{i,i+1}$ that updates    the bond $(i,i+1)$ located
   in the bulk of the system  can be  expressed
   with the help of the Pauli matrices. 
 We recall that  Pauli matrices    are given  by
\begin{equation} 
  S^x =  \left( \begin{array}{cc}
           0 & 1  \\ 1 & 0 \end{array} \right) \,   ,  \,\,\,\,\,
 \,\,\,\,\,  S^y =  \left( \begin{array}{cc}
           0 & -i  \\ i & 0 \end{array} \right) \, , \,\,\,\,\, 
\,\,\,\,\,   S^z =  \left( \begin{array}{cc}
           1 & 0  \\ 0 & -1 \end{array} \right)   \,\,\,\,\,.  
  \end{equation}
 We  also need  the following  operators
\begin{equation} 
 S^{+} = \frac{ S^x + i S^y}{2} =  \left( \begin{array}{cc}
           0 & 1  \\ 0 & 0 \end{array} \right) \, ,   \,\,\,\,\,
 S^{-}  = \frac{ S^x - i S^y}{2} =  \left( \begin{array}{cc}
           0 & 0  \\ 1 & 0 \end{array} \right)   \,\,\,\,\,. 
 \end{equation}
  After identifying  the spin-$\frac{1}{2}$ basis  
  $(|\uparrow\rangle,  |\downarrow\rangle)$  of  ${\mathbb  C}^2$
   with  the local basis
  $(|1\rangle_i, |0\rangle_i)$   of the two-dimensional space
   ${\mathbb  C}^2_i$ associated with the site $i$, 
   we  define an action of  the Pauli matrices  on the ASEP 
 configuration space  $\mathcal{S}$ as follows~: 
\begin{equation} 
   S^{a}_i =   {\bf 1}_1 \otimes  \ldots  \otimes {\bf 1}_{i-1}  \otimes
     S^{a} \otimes  {\bf 1}_{i+1}  \otimes  \ldots  {\bf 1}_{L}   
 \,\,\, \hbox{ with } \,\,\, a=x,y,z,+,- \, . \\
  \end{equation}
We observe that the local update operator  $M_{i,i+1}$
 can be written as 
\begin{equation} 
     M_{i,i+1} = p  S^{-}_i S^{+}_{i+1}  + q  S^{+}_i S^{-}_{i+1}
 +  \frac{1}{4}  S^z_i S^z_{i+1} - \frac{p-q}{4} ( S^z_i -  S^z_{i+1})
 -  \frac{1}{4}  \, , 
\label{eq:localXXY}
\end{equation}
where $ {\bf 1}$ represents the identity operator. 
 The Markov matrix of the ASEP on the periodic lattice of size $L$
 is thus given by 
 \begin{equation} 
   M =  \sum_{i=1}^L \left( p  S^{-}_i S^{+}_{i+1}  + q  S^{+}_i S^{-}_{i+1}
   +  \frac{1}{4}  S^z_i S^z_{i+1} \right) - \frac{L}{4}  \, , 
\label{eq:XXY}
\end{equation}
 (we recall that the site $L+1$ is the same as   the site number 1).
 The Markov matrix is therefore expressed as  a  spin chain
 on the periodic lattice. This spin chain is  non-Hermitian
 when $p \neq  q$. For  $ p = q = 1/2$, the  Markov matrix $M$
 is identical  with  the antiferromagnetic Heisenberg XXX spin chain 
 which was exactly solved by Hans  Bethe (1931).  More generally, 
 a  similarity transformation allows us to map exactly 
 the  Markov matrix $M$ on an antiferromagnetic XXY 
quantum spin chain with   hermiticity
 breaking   boundary terms. Following Essler and Rittenberg (1996), 
 we  consider the operators 
\begin{equation} 
  U_i =  {\bf 1}_1 \otimes  \ldots  \otimes {\bf 1}_{i-1}  \otimes
    \left( \begin{array}{cc}
           1 & 0  \\ 0 & \alpha^{i-1} \end{array} \right)
    \otimes  {\bf 1}_{i+1}  \otimes  \ldots  {\bf 1}_{L}          
       \label{def:Ui}  \, , 
\end{equation}
 where we have defined
   \begin{equation} 
       \alpha  = \sqrt{ \frac{p}{q}}
          \label{def:alpha}  \, . 
\end{equation}
 The operator $H$  defined as
  \begin{equation} 
       H =   (\alpha + \alpha^{-1})   \left(\prod_{i=1}^L  U_i \right)
 \left( M  +  \frac{L}{4} {\bf 1} \right)
    \left(\prod_{i=1}^L  U_i \right)^{-1}  \, , 
 \end{equation}
  is  then  given by 
\begin{equation} 
 H   = \frac{1}{2}
\sum_{i=1}^{L-1} \left( S^{x}_i S^{x}_{i+1}  +   S^{y}_i S^{y}_{i+1}
 +  \frac{\alpha + \alpha^{-1}}{2}  S^{z}_i S^{z}_{i+1} \right)
  +  \alpha^L S^{-}_L S^{+}_{1}
 +  \alpha^{-L} S^{+}_L S^{-}_{1}
 +  \frac{\alpha + \alpha^{-1}}{4}  S^z_L S^z_{1} \, . 
\label{eq:XXY2}
\end{equation}
 In this representation, we can formally  rewrite $H$ as  
  the XXY spin chain  on
 a  periodic lattice   with  the following {\it  twisted boundary conditions} 
\begin{eqnarray}
         S^{+}_{L+1} &=&   \alpha^L  S^{+}_{1}   \nonumber \\ 
         S^{-}_{L+1} &=&   \alpha^{-L}  S^{-}_{1}   \nonumber \\ 
          S^z_{L+1} &=&     S^z_{1} \, .  
\end{eqnarray}  

\subsection{The Yang-Baxter equation}

 We shall now derive  the  Yang-Baxter  equation which  
 will allow us to define  in  the subsection \ref{sub:Transfer}
 a one-parameter family of commuting operators
 that contains  the Markov matrix $M$. Such a family of  
commuting operators ensures the existence of a sufficient  number
 of conserved quantities that fully  label the states of the  ASEP.
 This property  is the  key to the  integrability of the ASEP.

 Let us  consider two  {\it auxiliary}   sites  labeled as 
 $a$ and $b$,  each of them  having  two possible  states    
   $(|1\rangle_a, |0\rangle_a)$  and  $(|1\rangle_b, |0\rangle_b)$ 
 respectively. The four possible states of  $a$ and $b$
 span a  four  dimensional complex
  vector space  denoted as  ${\mathcal A } \otimes {\mathcal B}$. For any
 given  value of the complex number $\lambda$ (called the
 {\it  spectral parameter}), we   define an      operator 
  $ {\mathcal L}_{ab}(\lambda)$ which acts on this  tensor space 
 as follows
\begin{equation} 
     {\mathcal L}_{ab}(\lambda) = P_{ab} \left( 1 + \lambda M_{ab}\right) \, ,
 \label{eq:defLi}
 \end{equation}
 where  $M_{ab}$ and $P_{ab}$ represent the jump and the  permutation 
 operators  between   $a$ and $b$  as defined in equations~(\ref{eq:localM})
 and (\ref{eq:localP}) respectively. We emphasize that the auxiliary  sites
  $a$ and $b$  should not be viewed as neighbouring sites on a given lattice
  but rather as  an `abstract'  pair of 
 sites related by  non-local  jump and exchange 
  operators   $M_{ab}$ and $P_{ab}$. We 
  now consider three auxiliary sites  $a$,  $b$ and $c$, and 
prove   the  Yang-Baxter  relation~:
\begin{equation} 
   {\mathcal L}_{ab}(\nu) {\mathcal L}_{cb}(\lambda)   
 {\mathcal L}_{ca}(\mu) = {\mathcal L}_{ca}(\mu)
  {\mathcal L}_{cb}(\lambda) {\mathcal L}_{ab}(\nu) 
\,\,\, \hbox{ for } \,\,\, 
  \nu =  \frac{\lambda - \mu}{ 1 - \mu +  pq\lambda \mu } \, .
 \label{eq:YB1}
 \end{equation}

   We shall need the fact   that 
 the      jump and   permutation  operators between the auxiliary 
 sites  satisfy  the following
 relations analogous  to~(\ref{eq:PM}--\ref{eq:MiMj}). For
  example, we have 
\begin{eqnarray}
     P_{c,a} M_{a,b}  &=&  M_{c,b}  P_{c,a}  \,  \label{eq:PM2}  \\
          M_{a,b}^2  &=&       -  M_{a,b} \,  \label{eq:M2M2} \\
     M_{c,a}\,  M_{a,b} \,  M_{c,a} &=& pq  M_{c,a}\,   
      \label{eq:MMM2} \,  \\
      M_{a,b}\,  M_{c,a} \,  M_{a,b} &=& pq  M_{a,b}\,   \, .
      \label{eq:MMM2b}  
\end{eqnarray}
  In order to derive the  Yang-Baxter    equation, we first remark that 
\begin{eqnarray}
  {\mathcal L}_{ab}(\nu) {\mathcal L}_{cb}(\lambda)   
 {\mathcal L}_{ca}(\mu) &=&   P_{ab} \left( 1 + \nu M_{ab}\right)
     P_{cb} \left( 1 + \lambda M_{cb}\right)
   P_{ca} \left( 1 + \mu  M_{ca}\right) \nonumber \\
    &=&   P_{ab} \left( 1 + \nu M_{ab}\right)   P_{cb}
    P_{ca} \left( 1 + \lambda M_{ab}\right)  \left( 1 + \mu  M_{ca}\right)
    \nonumber \\   &=&   P_{ab}  P_{cb}  P_{ca}
   \left( 1 + \nu M_{ca}\right) \left( 1 + \lambda M_{ab}\right)
\left( 1 + \mu  M_{ca}\right) \, .
\end{eqnarray}
 Similarly,  we have
 \begin{equation} 
 {\mathcal L}_{ca}(\mu)  {\mathcal L}_{cb}(\lambda) {\mathcal L}_{ab}(\nu) 
  =   P_{ca}  P_{cb}  P_{ab}
   \left( 1 + \mu M_{ab}\right) \left( 1 + \lambda M_{ca}\right)
\left( 1 + \nu  M_{ab}\right) \, .
  \end{equation}
 Now, with the help of the  Temperley-Lieb algebra,
 equations~(\ref{eq:PM}--\ref{eq:MiMj}),
 we obtain  
\begin{eqnarray}
&&  \left( 1 + \nu M_{ca}\right) \left( 1 + \lambda M_{ab} \right)
\left( 1 + \mu  M_{ca}\right)  \nonumber \\
  &=&  1 + (\mu + \nu)  M_{ca}
 +  \lambda M_{ab} + \mu\nu  M_{ca}^2 +  \lambda\mu M_{ab} M_{ca}
 + \lambda\nu M_{ca} M_{ab} +  \lambda\mu\nu M_{ca} M_{ab} M_{ca}
   \nonumber \\   &=& 1 +  (\mu + \nu - \mu\nu +pq  \lambda\mu\nu  )  M_{ca}
  +  \lambda M_{ab} +   \lambda\mu M_{ab} M_{ca}  + 
 \lambda\nu M_{ca} M_{ab} \,.
\end{eqnarray}
 Similarly we have 
\begin{eqnarray} 
&& \left( 1 + \mu M_{ab}\right) \left( 1 + \lambda M_{ca}\right)
\left( 1 + \nu  M_{ab}\right) \nonumber \\  &=&  1 +  
 (\mu + \nu - \mu\nu +pq  \lambda\mu\nu  )  M_{ab} +   \lambda  M_{ca}
 + \lambda\mu M_{ab} M_{ca}  +  \lambda\nu M_{ca} M_{ab} \,. 
  \end{eqnarray}
 Using the fact that  $  P_{ab}  P_{cb}  P_{ca} =  P_{ca}  P_{cb}  P_{ab}$
 we  complete  the proof of equation~(\ref{eq:YB1}). 

\subsection{The Transfer Matrix}
 \label{sub:Transfer}

 We are now ready to apply the Algebraic Bethe Ansatz to the ASEP
(for  introduction  to this subject, see Faddeev 1984, de Vega 1989, 
 and Nepomechie 1999). We introduce an auxiliary site
  (denoted as site 0) which can be in two states 
   $ |1\rangle_0$ or   $ |0\rangle_0$. These
 two states  span  a two dimensional vector space, ${\mathcal A}$,
 the auxiliary space. We define, as in equation~(\ref{eq:defLi}), 
 a  local  transfer  operator    $ {\mathcal L}_{i0}(\lambda)$ 
 between the site $i$ of the ASEP ring and the auxiliary site 0. 
This operator,  that we shall denote as  $ {\mathcal L}_{i}(\lambda)$,  
  can   be  represented   as  a  $2 \times 2$
  operator  on   the vector space ${\mathcal A}$, 
\begin{equation} 
     {\mathcal L}_{i}(\lambda)  =   \left( \begin{array}{cc}
                   a(\lambda) & b(\lambda)  \\
                   c(\lambda) & d(\lambda)  
                         \end{array}
                    \right)     \, , 
\label{eq:defLi2}
 \end{equation}
   where  the  matrix elements
  $a(\lambda),  b(\lambda),  c(\lambda)$  and $d(\lambda)$ 
  are themselves  $2^L \times 2^L$  operators
 that act    on  the configuration space  ${\mathcal S}$.
  These operators    act trivially on all sites
 different from $i$ and  are given by 
\begin{eqnarray} 
 a(\lambda) &=& {\bf 1}_1 \otimes  \ldots {\bf 1}_{i-1}\otimes
             \left( \begin{array}{cc}  1 & 0 \\
                                       0 & q\lambda   \end{array} \right) 
      \otimes {\bf 1}_{i+1} \ldots {\bf 1}_{L} \, ,\label{eq:defaL} \\ 
b(\lambda)  &=& {\bf 1}_1 \otimes \ldots {\bf 1}_{i-1}\otimes
              \left( \begin{array}{cc}  0 & 0  \\
                                      1 - p\lambda & 0   \end{array} \right)
 \otimes {\bf 1}_{i+1} \ldots {\bf 1}_{L}  \, ,\label{eq:defbL} \\
 c(\lambda)  &=& {\bf 1}_1 \otimes \ldots {\bf 1}_{i-1}\otimes
                       \left( \begin{array}{cc}  0 & 1 - q\lambda \\
                                       0 & 0   \end{array} \right) 
   \otimes {\bf 1}_{i+1} \ldots {\bf 1}_{L}  \, ,\label{eq:defcL} \\
d(\lambda)  &=& {\bf 1}_1 \otimes \ldots {\bf 1}_{i-1}\otimes
           \left( \begin{array}{cc}  p\lambda   & 0 \\
                                   0 &       1  \end{array} \right) 
      \otimes {\bf 1}_{i+1} \ldots {\bf 1}_{L} \label{eq:defdL}  \, . 
\end{eqnarray}

 We  now   consider the operator
\begin{equation}
  {\mathcal R}(\nu) = 1 + \nu M_{a', a}  \, ,
\end{equation}  
  that acts   on  ${\mathcal A} \otimes {\mathcal A}'$
  where  the  auxiliary
 spaces ${\mathcal A}$ and ${\mathcal A}'$  are the configuration spaces of
  the  auxiliary sites $a$ and $a'$.   
 In   the basis   $ \left( |1_a, 1_{a'}\rangle,|1_a, 0_{a'}\rangle,
   |0_a, 1_{a'}\rangle, |0_a, 0_{a'}\rangle \right)$ of 
  ${\mathcal A} \otimes {\mathcal A}'$, the 
  operator ${\mathcal R}(\nu)$   is  
 represented by a $4 \times 4$ scalar  matrix~: 
\begin{equation}
 {\mathcal R}(\nu)  =    \left( \begin{array}{cccc}
  1 & 0 & 0 & 0 \\
  0 & 1- q\nu & p\nu  & 0 \\
  0 &  q\nu & 1 - p\nu & 0 \\
  0 & 0 & 0 & 1
        \end{array} \right)     \, .     
\label{eq:MatR}
\end{equation}
 
 Using equation~(\ref{eq:YB1}),
 we  find  that    the operators  $ {\mathcal L}_{i}(\lambda)$  satisfy the 
   Yang-Baxter  equation~: 
\begin{equation}
 {\mathcal R}(\nu)
   \big[ {\mathcal L}_{i}(\lambda) \otimes {\mathcal L'}_{i}(\mu)
    \big]
=   \big[  {\mathcal L}_{i}(\mu) \otimes {\mathcal L'}_{i}(\lambda)  \big]
    {\mathcal R}(\nu) \,\,\, {\rm with }  \,\,\,
  \nu =  \frac{\lambda - \mu}{ 1 - \mu + pq\lambda\mu } \, , 
\label{eq:YBE}
\end{equation}
 where  ${\mathcal L}_{i}$ and ${\mathcal L'}_{i}$  
 are interpreted as  $2 \times 2$  matrices
  acting, respectively,  on  ${\mathcal A} $  and ${\mathcal A'}$ 
  with  elements that are themselves operators on ${\mathcal S}$.
 Their tensor product is thus a $4 \times 4$  matrix,
 acting  on  ${\mathcal A}\otimes {\mathcal A'}$ 
 with  matrix elements that are operators on ${\mathcal S}$.

 The  {\it monodromy}  matrix is defined as
\begin{equation}
  T(\lambda) =  {\mathcal L}_{1}(\lambda) {\mathcal L}_{2}(\lambda)
 \ldots  {\mathcal L}_{L}(\lambda) 
 \, ,
\label{eq:defTlambda}
\end{equation}
 where the product of the  ${\mathcal L}_{i}$'s has to be understood as
  a product of  $2 \times 2$ matrices acting on ${\mathcal A}$
 with non-commutative elements. 
 The  monodromy  matrix  $T(\lambda)$
  can thus  be written as
\begin{equation}
  T(\lambda) =  \left( \begin{array}{cc}
                      A(\lambda)  &  B (\lambda)  \\
                      C(\lambda)  &  D (\lambda) \end{array}
                    \right)     \, ,
\label{eq:def2T}
\end{equation}
where  $A, B, C$ and $D$ are operators on the configuration 
  space  ${\mathcal S}$.  By applying  
  equation~(\ref{eq:YBE})  to  all the lattice sites $i$,
 we  find  that the monodromy matrix satisfies a similar relation
\begin{equation}
 {\mathcal R}(\nu)
   \big[ T(\lambda) \otimes T(\mu)
    \big]
=   \big[ T(\mu) \otimes T(\lambda)  \big]
    {\mathcal R}(\nu) \,\,\, {\rm with }  \,\,\,
  \nu =  \frac{\lambda - \mu}{ 1 - \mu + pq\lambda\mu } \, , 
\label{eq:YBE2}
\end{equation}

 Taking the trace of the   monodromy  matrix
 over the auxiliary space  $\mathcal{A}$, we obtain a one-parameter
 family of {\it  transfer}  matrices   acting on  ${\mathcal S}$
\begin{equation}
  t(\lambda) = 
{\rm Tr}_\mathcal{A}( T(\lambda)) =   A(\lambda) +  D (\lambda) \, .
  \label{eq:deftlam}
\end{equation}
 By taking the trace   of 
   equation~(\ref{eq:YBE2})  over ${\mathcal A}\otimes {\mathcal A'}$
  and using the fact that
 ${\mathcal R}(\nu)$ is generically an invertible matrix,
 we deduce that   the operators
  $t(\lambda)$   form a family of commuting operators
 (see e.g.,  Nepomechie, 1999). 
  In particular,  this family contains the  translation  operator
  ${ T} =  t(0)$ 
 (that shifts  all the particles   simultaneously 
  one site forward) and 
 the Markov matrix   $M = t'(0)/t(0)$.

  The equation~(\ref{eq:YBE2}), when  written explicitly, leads
 to 16 quadratic  relations between the operators
  $A(\lambda)$, $B(\lambda)$, $C(\lambda)$, $D(\lambda)$ and the operators
 $A(\mu), B(\mu), C(\mu), D(\mu)$. In particular,  we have
\begin{eqnarray}
        C(\lambda) C(\mu)  &=&  C(\mu) C(\lambda)        \label{CC} \, ,\\
        A(\lambda) C(\mu)  &=&   ( 1 - q\nu )  A(\mu) C(\lambda)
     + p\nu  C(\mu)     A(\lambda)   \label{AC} \, ,\\
     D(\lambda)  C(\mu)    &=&  p\nu C(\mu) D(\lambda) +  ( 1 - p\nu)
       D(\mu) C(\lambda)    \label{DC} \,  .
\end{eqnarray}
 These relations are used in the next subsection to construct
 the eigenvectors of the family of transfer matrices  $t(\lambda)$.

 \subsection{The Bethe equations}

  Using  Algebraic Bethe Ansatz, 
 the  common eigenvectors of this    family  are   explicitly  constructed 
  by actions  of  the  $C$ operators on
  the  reference   state $\Omega$, defined as
 \begin{equation}
       \Omega =  | 0_10_2 \ldots 0_L \rangle  \, . 
   \label{eq:defOmega}
\end{equation}
 The state   $\Omega$  corresponds to a  configuration where all
  the sites are empty.  The operators   $A(\lambda)$ and $D(\lambda)$
 have a simple action on  $\Omega$, namely
\begin{equation}
 A\left(\frac{\lambda}{q}\right)\, \Omega  = \lambda^L \,  \Omega
 \,\,\,\, \hbox{ and }
  \,\,\,\,  D\left(\frac{\lambda}{q}\right)\, \Omega  = \, \Omega 
  \label{Aomeg} \, .
\end{equation}
   Now,  for any  $N \le L$,   we define the vector
\begin{equation}
    | z_1, z_2 \ldots  z_N    \rangle  
 =     C\left(\frac{z_N}{q}\right) \ldots C \left(\frac{z_2}{q}\right)
  C \left(\frac{z_1}{q}\right)\,  \Omega \,,  
\label{eq:ABA}
\end{equation}
 where $ z_1, z_2 \ldots  z_N$ are complex numbers. 
  Each operator $C$ creates a particle  in the system, and 
 the   state $| z_1, z_2 \ldots  z_N   \rangle$
  is a linear combination  of
 configurations  with exactly  $N$ particles. This vector   
 is an eigenvector of the  operator  $t(\lambda/q)$ 
 (for all values of $\lambda$) 
   and in particular
 of the Markov matrix M,  provided  the pseudo-momenta 
  $z_1$, $z_2$, $\ldots$,
   $z_N$ satisfy the Bethe equations (Gwa and Spohn 1992)~:
\begin{equation}
   z_l^L  = (-1)^{N-1} \prod_{i =1}^N \frac{p z_l z_i - z_l + q}
  {p z_l z_i -z_i +q } \,\,\,
  {\rm  for }\,\,\,\,  l =1 \ldots N\, . 
\label{eq:Bethepq}
\end{equation}
These equations are derived by considering the action of the
 operators  $t(\lambda/q)$   on 
$ | z_1, z_2 \ldots  z_N    \rangle$~:
\begin{equation} 
t\left(\frac{\lambda}{q}\right)  \,  | z_1, z_2 \ldots  z_N \rangle =
 A\left(\frac{\lambda}{q}\right)  \, 
  C\left(\frac{z_N}{q}\right) \ldots C \left(\frac{z_2}{q}\right)
  C \left(\frac{z_1}{q}\right)\, \Omega
 +  D\left(\frac{\lambda}{q}\right)   \, 
C\left(\frac{z_N}{q}\right) \ldots C \left(\frac{z_2}{q}\right)
  C \left(\frac{z_1}{q}\right)\, \Omega  \label{tOmega} \, .
 \end{equation} 
 Applying repeatedly
 equations~(\ref{AC}) and~(\ref{DC})  to the first term  and
 to the second term on the r.h.s. of  this equation, respectively,
   we obtain   a linear
 combination  of  $ | z_1, z_2 \ldots  z_N \rangle$
and of  $N$  vectors  $ | z_1, \ldots  z_{i-1},\lambda, z_{i+1}
  z_N \rangle$ for $i=1,\ldots,N$. If we want 
  $| z_1, z_2 \ldots  z_N   \rangle$ to be an eigenvector
  of $t(\lambda/q)$, the  coefficients of these  $N$  vectors  must
 vanish . The vanishing conditions for  the
 coefficients of the   unwanted terms provide 
 a set of  $N$ relations that must be satisfied
  by  the  complex numbers  $ z_1, z_2 \ldots  z_N$.  These relations, 
  when  written explicitly,  are precisely 
 the Bethe equations~(\ref{eq:Bethepq}). 

The corresponding eigenvalue of  $t(\lambda/q)$ is given by 
 \begin{equation}
     \mathcal{E}(\lambda; z_1, z_2 \ldots  z_N ) = 
    \prod_{i =1}^N \frac{p \lambda z_i -\lambda  +q }{z_i -\lambda}  
  +   \lambda^L   \prod_{i =1}^N \frac{p \lambda z_i -z_i +q}{\lambda - z_i} 
  \,.
 \label{eq:eigenvalpq}
 \end{equation}
 Using the Bethe equations~(\ref{eq:Bethepq}),  we find 
 that  $\mathcal{E}(\lambda)$ is a polynomial in  $\lambda$
 of degree $L-N$. The eigenvalue of the Markov matrix
 is given by 
 \begin{equation}
     E(z_1, z_2 \ldots  z_N ) = p \sum_{i =1}^N z_i + 
          q \sum_{i =1}^N \frac{1}{ z_i} - N \, .
 \label{eq:Meigenvalpq}
 \end{equation}

  The Bethe  equations~(\ref{eq:Bethepq}) for the ASEP were first derived and
 analyzed by  Gwa and Spohn (1992), who used
  the coordinate Bethe Ansatz,  a method  more elementary 
 and quicker than the one presented here.   We have chosen  to
 describe here  the more sophisticated Algebraic Bethe Ansatz technique,
 because, in our opinion,  the ASEP provides  the simplest
 pedagogical example of  this method.  Besides, the  
 Algebraic Bethe Ansatz becomes  unavoidable to study 
 exclusion processes 
 with different classes of particles and   systems with
 open boundaries (de Gier and Essler 2005). Finally, 
 the  Algebraic Bethe Ansatz allows to construct 
 a Matrix Product Ansatz involving  quadratic algebras and
 to draw a rigorous relation between the two complementary
 approaches used  to  analyze   the ASEP (Golinelli
 and Mallick 2006).

 \section{Analysis of the Bethe equations of the TASEP}
 \label{sec:eqTASEP}

 In the limit of a totally asymmetric exclusion process
  the Bethe equations become particularly simple
 and can be analyzed thoroughly. In this section we 
 present some exact results that are valid for the TASEP.
  Substituting  $p=0$ and $q =1$ in the  Bethe equations~(\ref{eq:Bethepq}),
 we obtain 
\begin{equation}
   z_i^L  = (-1)^{N-1} \prod_{j =1}^N \frac{1 -z_i}{1 -z_j} \,\,\,
  {\rm  for }\,\,\,\,  i =1 \ldots N\, .
\label{eq:Bethe}
\end{equation} 
In terms of  the  variables $Z_i = 2/z_i -1$, these equations
 become  (Gwa and Spohn 1992)
 \begin{equation}
  (1-Z_i)^N \ (1+Z_i)^{L-N}  
  =  - 2^L \prod_{j=1}^N \frac{Z_j - 1}{Z_j + 1}  \ \ \
  \hbox{for}    \ \ \ i=1,\dots,N     \, .
  \label{eq:bez}   
\end{equation} 
 We note that the right-hand side of these equations is independent
 of the index $i$: this property is true only for the TASEP
 where  the  Bethe equations decouple and can be  reduced
 to an effective one-variable equation.
 The corresponding eigenvalue $E$ of  the Markov matrix  $M$ is given by 
 \begin{equation}
 2  E =     -N + \sum_{j=1}^N Z_j  \, .
 \label{eq:eigenval}
 \end{equation}
 For the TASEP, it is  possible to show that the associated 
 Bethe eigenvector $P_E$  is a determinant (Golinelli and Mallick 2005b)
\begin{equation}
   P_E (x_1,\dots,x_N) = \det(R)  \, , 
   \label{eq:psidet}
\end{equation}
where $R$ is a $N \times N$ matrix with elements
\begin{equation}
   R(i,j) = \frac{z_{i}^{x_j}}{(1-z_i)^j} \ \ \mbox{for } 1 \le i,j \le N  \, ,
   \label{eq:r}
\end{equation}
 where $(z_1, \dots, z_N)$ are the roots of the
  Bethe equations~(\ref{eq:Bethe}). Similar formulae with  determinants 
  also appear in the expression of $P_t$ calculated
 for an arbitrary initial condition $P_0$ on an infinite
 open lattice  (Sch\"utz 1997, Sasamoto and Wadati 1998,
  R\'akos and Sch\"utz 2005)  and on a periodic lattice (Priezzhev 2003).

\subsection{Layout of  the solutions of the Bethe equations}
  
 The   solutions of the   Bethe  equations~(\ref{eq:bez})
 are the roots  of the  polynomial equation of  degree $L$ 
\begin{equation}
   (1-Z)^N (1+Z)^{L-N} = Y \, , 
  \label{eq:zzy}
\end{equation}
 where  $Y$ must be  determined self-consistently
 by the r.h.s. of  equation~(\ref{eq:bez}).  In this subsection,
 we describe the geometrical layout of the roots 
 of the  polynomial~(\ref{eq:zzy}) for  an arbitrary 
 value of the  complex number $Y$.

 The $L$ solutions $(Z_1,\dots, Z_L)$ 
  of this equation display interesting geometric properties.
 We explain how to label these  solutions so 
 that each $Z_k(Y)$ becomes  an analytic
function of the parameter $Y$ in the complex plane with a branch cut
along the real semi-axis $[0,+\infty)$.

\begin{figure}[th]
\centerline{\includegraphics*[width=0.70\textwidth, angle =0]{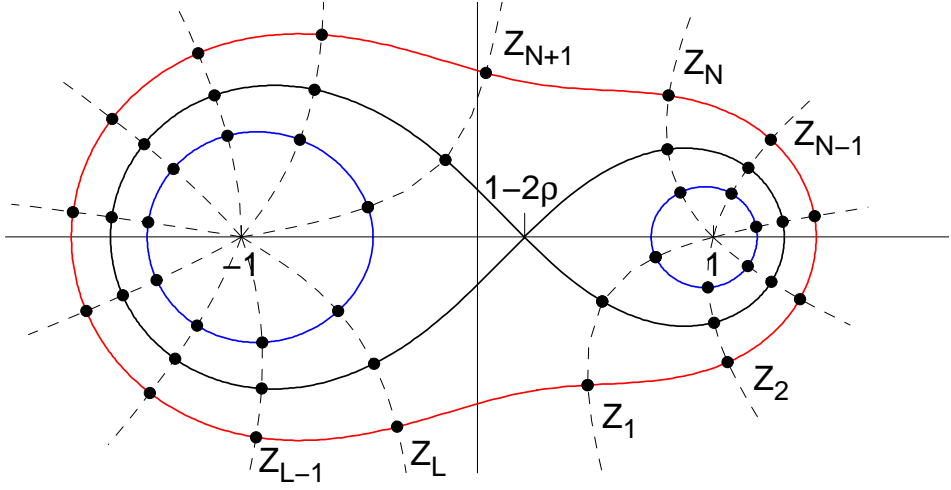}}
  \caption{\em 
    Labelling the roots of the Bethe equations.  Here $L = 15$, $N=
    6$, $\phi = \pi/2$ and $r/r_c = 0.8,1, 1.2$ (see text).  The
    continuous curves are  Cassini ovals.  When $r$ is fixed and $\phi$ varies
    from 0 to $2\pi$, each $Z_k$ slips counterclockwise along a part
    of the Cassini oval.  When $\phi$ is fixed and $r$ varies from
    $\infty$ to 0, each $Z_k$ travels along a dashed curve from
    $\infty$ to points +1 or -1.  }
  \label{fig:cassini}
\end{figure}

A non-zero complex number $Y$ can be written in a unique way as
\begin{equation}
   Y = r^L \, e^{i \phi} 
   \ \ \mbox{with}\ \  
   0 \le \phi < 2 \pi
   \ ,
\end{equation}
$r$ being a positive real number.  We emphasize that  the argument
has a branch cut along $[0,+\infty)$~:  $\phi$ has a discontinuity
 of $2\pi$ when $Y$ crosses the positive real axis.
  For a given value of $r$, the
complex numbers $Z_k$ belong to the generalised Cassini oval defined
by
\begin{equation}
   |Z-1|^{\rho} |Z+1|^{1-\rho} = r
\end{equation}
where $\rho = N/L$ is the filling of the system.  As shown in
 Fig.~\ref{fig:cassini}, the topology of the Cassini ovals  depends on
the value of $r$ with a critical value
\begin{equation}
   r_c = 2 \rho^\rho(1-\rho)^{1-\rho} \, ;
   \label{eq:rc}
\end{equation}
\begin{itemize}
\item for $r < r_c$, the curve consists of two disjoint ovals with $N$
  solutions on the oval surrounding $+1$ and $L-N$ solutions on the
  oval surrounding $-1$.
  
\item for $r = r_c$, the curve is a deformed Bernouilli lemniscate
  with a double point at $Z_c = 1 - 2 \rho$.
  
\item for $r > r_c$, the curve is a single loop with $L$ solutions.
\end{itemize}
The Cassini ovals are symmetrical only if $\rho = 1/2$.

In order to label the solutions, we start by considering the limit $r
\to \infty$ for a given $\phi$.  Equation~(\ref{eq:zzy}) then becomes
\begin{equation}
  Z^L \sim r^L \, \exp [i(\phi - N\pi)]  \, .  
\end{equation}
The solutions $Z_k$ are given  by
\begin{equation}
   Z_k \sim r \, \exp\left[ \frac{i}{L} [\phi - N\pi + 2(k-1)\pi] \right]
   \ \ \mbox{with}\ \  
    k = 1, \ldots,  L
   \, .
   \label{eq:zk}
\end{equation}
In other words, the $Z_k$ are regularly distributed along a large
circle of radius $r$ with
\begin{equation}
   \frac{\phi - N\pi}{L} \le \arg Z_1 < \dots < \arg Z_L < 
    \frac{\phi - N\pi}{L} + 2\pi
    \, .
\end{equation}

This labelling, obtained for large $r$, is extended by analytic
continuation to all values of $r$, keeping $\phi$ {\em fixed}.  The
loci of the $Z_k$ are plotted  in Fig.~\ref{fig:cassini} (dashed curves):
they are orthogonal to the Cassini ovals.  A singularity appears along
the branch $\phi = 0$ because $Z_1$ and $Z_{N+1}$ collapse into each
other at the double point $Z_c$ when $r=r_c$; we circumvent it by
choosing $\phi = 0^+$.

With this labelling, the solutions are ordered along the Cassini
ovals.  Moreover, when $r < r_c$, the solutions $(Z_1,\dots,Z_N)$
group together on the right oval and $(Z_{N+1}, \dots, Z_L)$ on the
left oval.

  \subsection{Procedure for solving the Bethe Equations}
 \label{sec:procedure}

 The special `one-body' structure of the
  TASEP  Bethe equations leads to the following self-consistent
 method for solving them. In the following sections, we shall
 show how this procedure can  effectively be used to calculate
 the spectral gap of the model and to  determine some arithmetical
 global properties of the spectrum. 

\begin{itemize}
\item { SOLVE},  for any given value of $Y$,  the polynomial equation 
        $ \,\,\, ( 1 -Z_i)^N   ( 1 + Z_i)^{L-N}  =  Y \, . $
   The  $L$ roots  $Z_1(Y)\,  \ldots \,  Z_L(Y)$ of this equation 
are located on   a  {\it Cassini Oval} 
\item   {\it CHOOSE   $N$ roots  }
  $Z_{c(1)}(Y), \ldots \, Z_{c(N)}(Y)$ amongst  the $L$ available roots,
 {\it i.e.}, select  a  {\it choice set} $c : \{ c(1),\ldots , c(N)\} \subset
 \{ 1,\ldots ,L\}  \, . $
\item   { SOLVE }  the {\it self-consistent}
 equation ${   A_c(Y) = Y }$  where
          $$A_c(Y) =  -2^L \prod_{j=1}^N  
 \frac{ Z_{c(j)}(Y) -1}{ Z_{c(j)}(Y) + 1}  \, .$$
 The  self-consistent equation $  A_c(Y) = Y $ is a fixed point
 equation. It is solved in practice  by choosing  iteratively a new $Y'$
given by  $ Y'=   A_c(Y)$ and going back to the first step
 until convergence is reached.
\item  { CALCULATE }   from the  fixed point  value of $Y$,   the 
  $Z_{c(j)}(Y)$'s and the energy corresponding to the choice 
  set  $c$~:
       $$ 2E_c(Y) = -N + \sum_{j=1}^N Z_{c(j)}(Y) .$$
\end{itemize}

  Consider the choice function $c_0(j) = j$ that selects
  the $N$ fugacities $Z_i$ with
 the largest real parts. The   $A$-function and eigenvalue
  associated  with this  choice function   are given by 
\begin{eqnarray}
  A_0(Y) &=& -2^L \prod_{j=1}^N \frac{Z_j - 1}{Z_j + 1} \, , \label{eq:a0} \\
  2 E_0 &=& -N + \sum_{j=1}^N Z_j  \, .\label{eq:e0}
\end{eqnarray}

 In (Golinelli and Mallick 2004a, 2005a), we derived the
 following  {\it  exact}  combinatorial
 formulae for $A_0(Y)$ and  $E_0(Y)$ for any finite values
 of $L$ and   $N$~:
\begin{eqnarray}
 \log \frac{A_0(Y)}{Y} &=&  \sum_{k=1}^\infty 
 \left(\begin{array}{c} kL\\kN \end{array} \right) 
 \frac{Y^k}{k2^{kL}}    \, ,  \label{eq:CombA} \\
  E_0(Y) &=& -   \sum_{k=1}^\infty 
   \left(\begin{array}{c} kL -2 \\kN -1 \end{array} \right) 
 \frac{Y^k}{k2^{kL}}       \,  .    \label{eq:CombE}       
\end{eqnarray}
 These expressions  are derived  as follows:
 when $Y \to  0$, the $N$ roots of  the  equation~(\ref{eq:zzy})
  with the  largest real parts, {\it i.e.}, $(Z_1, \dots, Z_N)$
 converge to +1 (whereas the $L -N$ other roots go to -1).
 Consider a contour $\gamma$, positively oriented,
  that encircles +1 such that
 for $Y$ small enough  $(Z_1, \dots, Z_N)$  are inside  $\gamma$
 whereas  $(Z_{N+1}, \dots, Z_L)$ are  outside $\gamma$. Let 
 $h(Z)$ be a function, analytic in a domain that contains
  the contour  $\gamma$;   from the residue theorem, we deduce:
  \begin{equation}
  \sum_{j=1}^N h(Z_j) = \frac{1}{2i\pi}\oint_{\gamma}
 \frac{ (1 -Z)^{N-1} (1 +Z)^{L-N-1} ( L - 2N - LZ)}
  { (1 -Z)^{N} (1 +Z)^{L-N}  - Y } h(Z)  {\rm d}Z \, .
\label{eq:residus}
\end{equation}
The functions  $ \log\left({A_0(Y)}/{Y}\right)$ and $ E_0(Y)$  
 are  expressed as such 
 contour integrals by choosing $h(Z) = \log(\frac{1+Z}{2})$
 and $h(Z) = Z -1$, respectively.  
  Equations~(\ref{eq:CombA}) and~(\ref{eq:CombE}) result from 
expanding for small values of $Y$ the denominator
  in  equation~(\ref{eq:residus}), 
 thanks to the formula $(A - Y)^{-1} = A^{-1}\sum_{k=0}^\infty Y^k/A^k$
 valid for $ |Y| <| A|$. 

  The expressions~(\ref{eq:CombA}) and~(\ref{eq:CombE})
  are  then analytically continued in 
 $ {\mathbb C} - [1,\infty)$. 
 In the  limit $ L \to \infty$ and  with $\rho$ fixed,
  we obtain  from   the Stirling formula
\begin{equation} 
 \ln  \frac {A_0(Y)}{Y}  
    \to \frac{1}{\sqrt{2\pi \rho(1 -\rho) L}} 
 \ \Li_{3/2} \left(\frac{Y}{r_c^L} \right)
  \, ,   \label{eq:lhsa1}
\end{equation}
 where $r_c$ was defined in equation~(\ref{eq:rc}) and the  
 {\em polylogarithm} function  $\Li_{s}$ of index $s$  is given by
\begin{equation}
  \Li_s(z) = \frac{z}{\Gamma(s)} \ \int_0^{\infty} \frac{t^{s-1} \
  dt}{e^t -z} = \sum_{k=1}^{\infty} \frac{z^k}{k^s}  \, .  
  \label{eq:Li}
\end{equation} 
 The function $\Li_{s}$ is defined by the first
 equality on   the whole
 complex plane with a branch cut along the real semi-axis $[1,+\infty)$;  
 the second  equality is   valid only for $|z| < 1$.
 Similarly, when  $L\to \infty $,     $E_0(Y)$ can be expressed in terms
 of  the   polylogarithm  function $\Li_{5/2}$.

 From equation~(\ref{eq:CombA}), we observe  that the equation
$A_0(Y) = Y$ has the    solution $Y=0$ that
  yields $Z_j = 1$ for $j \le N $. This solution leads to $ E_0 = 0$. 
 The choice function $c_0(j) = j$ thus 
 provides the ground state of the Markov
 matrix. This fact has been verified numerically for small
 size  systems  (Gwa and Spohn 1992, Golinelli and Mallick 2004a).

 \subsection{Calculation of the Gap}

 The spectral gap, given by the 
  first excited eigenvalue,  corresponds to the choice 
  $c_1(j) = j$ for $j = 1, \ldots, N-1$ and $c_1(N) = N+1$
  (Gwa and Spohn 1992). 
 The associated self-consistency  function and eigenvalue 
 are given by 
\begin{eqnarray}
 A_1(Y) &=& A_0(Y) \frac{Z_{N+1}-1}{Z_{N+1}+1} \ \frac{Z_N+1}{Z_N -1} \, ,   
  \label{eq:a1} \\
  2E_1 &=& 2E_0 + (Z_{N+1} -  Z_N)  \, . 
  \label{eq:e1e0}
\end{eqnarray}
 
 In the large $L$ limit, 
 Bethe equations for the gap  become  at  the leading order 
\begin{equation}
  \Li_{3/2}(-e^{u\pi}) = 2i\pi \left[ (u+i)^{1/2} -  (u-i)^{1/2} \right]  ,
\hbox{  where $u$  is defined as }  \frac{Y}{r_c^L}= -{\rm e}^{u\pi} \,.
  \label{eq:li32}
\end{equation}
 The  solution of this equation  is given by:
\begin{equation}
  u = 1.119 \, 068 \, 802 \, 804 \, 474 \dots 
\label{eq:solu}
\end{equation}
 This leads to  the eigenvalue corresponding to the
 first excited state~: 
\begin{equation}
   E_1  =
   -2 \sqrt{ \rho( 1 - \rho )} \frac{6.509189337\ldots}{L^{3/2}} 
      \pm \frac{ 2 i \pi (2 \rho -1)}{L}  \, .  
\end{equation}
 We observe that the first excited state  consists
 in pair of conjugate  complex numbers when 
 $\rho$  is different from 1/2. The real part of $E_1$ describes the relaxation
 towards the stationary state. The corresponding relaxation
 time scales as the size $L$ 
  of the system raised to the power 3/2; the dynamical
 exponent $z$  of the ASEP is thus given by $z =3/2$. This value
 agrees with the dynamical exponent of the
  one-dimensional  Kardar-Parisi-Zhang equation
  that belongs to the same universality class
 as ASEP (Halpin-Healy and Zhang 1995). The imaginary part of
  $E_1$  represents the  relaxation oscillations and scales as $L^{-1}$; these
 oscillations correspond to a kinematic wave that propagates with
 the group velocity $2 \rho -1$ (Majumdar, private communication). 
 The method described here can be extended to calculate the
 higher  excitations of the spectrum above the ground state
 by considering other choice sets $c$. However, the precise  correspondence
 between the choice set and the level of the excitation has not been
 studied at present, to our knowledge.

 \subsection{Spectral degeneracies due to 
 a hidden symmetry of the Bethe equations}

 Although  the steady state of the ASEP 
  and the lowest excitations  have been extensively studied, little effort
   has been devoted to investigate 
  global spectral  properties of  the  Markov matrix.  In this section, we 
  show that the  Bethe equations of the TASEP  possess  an invariance
 property under exchange of roots that implies  
  the existence of unexpected  multiplets in the spectrum
   of the TASEP  evolution operator. This hidden  
symmetry of the Bethe equations allows to predict combinatorial formulae
for the orders of degeneracies and the number of multiplets of a given
order of degeneracy. 

 A  numerical diagonalization of the Markov
 matrix  reveals a striking global property of the TASEP spectrum~:
 the existence of large   degeneracies of very special orders. 
  Degeneracies of orders 2, 6, 7, 20... appear in 
  the  TASEP  spectrum at half-filling  (Table  \ref{tab:1/2}).
  This property does not follow from  the natural
   symmetries of the exclusion process (discussed  in section~\ref{sec:sym})
 which  suggest that the spectrum  should be composed only 
 of singlets for impulsion $k = \pm 1$  and of  singlets 
and doublets for $ k \ne \pm 1$.   
  Degeneracies   of higher order have no
reason to appear and therefore should not exist generically;
 however, multiplets do exist  in spectrum of  the  TASEP  at half-filling 
  (Table  \ref{tab:1/2}) as well as in the   TASEP at arbitrary filling
  (Table \ref{tab:other}).

\begin{table}  \centering
  \begin{tabular}{rr|rrrrr}
   $L$   & $N$ &  $m(1)$ &  $m(2)$ & $m(6)$ & $m(20)$ & $m(70)$
\\ \hline
    2    &     1     &     2 \\
    4    &     2     &     4  &     1  \\
    6    &     3    &     8  &     6  \\
    8    &     4    &    16  &    24  &    1 \\
   10    &     5    &    32  &    80  &   10 \\
   12    &     6    &    64  &   240  &   60  & 1      &     \\
   14    &     7    &   128  &   672  &  280  & 14     &     \\
   16    &     8    &   256  &  1792  & 1120  & 112    &  1  \\
   18    &     9    &   512  &  4608  & 4032  & 672    & 18  \\
  \end{tabular}
  \caption{\em 
    Spectral degeneracies in the TASEP at filling $\rho = 1/2$: 
  $L$ is the  size of the lattice, $N$ the number of particles, 
 $m(d)$   the number of  multiplets with degeneracy $d$.
   }
  \label{tab:1/2}
\end{table}

\begin{table}  \centering
  \begin{tabular}{r|r r|rrrrrr}
   $\rho$ & $L$ & $N$ & $m(1)$ & $m(2)$ & $m(3)$ & $m(4)$ & $m(5)$ & $m(15)$
\\ \hline
   1/3    &  9  &  3  &     81 &        &     1   \\
          &  12 &  4  &    459 &        &    12   \\
          &  15 &  5  &   2673 &        &    90  &     15   \\
          &  18 &  6  &  15849 &        &   540  &    270 &        &   1 \\
          &  21 &  7  &  95175 &        &  2835  &   2835 &    189 &  21 \\
   \hline 
   1/4    &  16 &  4  &   1816 &        &        &      1 \\
          &  20 &  5  &  15424 &        &        &     20 \\
          &  24 &  6  & 133456 &        &        &    240 &     36  \\
   \hline 
   1/5    &  25 &  5  &  53125 &        &        &        &      1  \\
   2/5    &  15 &  6  &   4975 &     15 \\
  \end{tabular}
  \caption{\em 
   Examples of spectral degeneracies in the TASEP at filling
   $\rho \ne 1/2$: 
  $m(d)$  is  the number of  multiplets with degeneracy $d$.
   }
  \label{tab:other}
\end{table}

These spectral degeneracies have an arithmetical origin and appear only
 when  $L$ (the number of sites) and $N$ (the number of particles) 
 are not relatively prime. We  thus define 
\begin{equation}
   \delta  = \gcd(L,N)
  \, .  \label{eq:defdelta}
\end{equation}
 The $L$  Bethe roots  of the polynomial (\ref{eq:zzy}) 
 can be grouped into   ${\delta}$  disjoint 
 {\it packages},  each   of  cardinality ${L/\delta}$. 
   The  roots composing  the 
  package ${\mathcal P}_s$  have the indices
  $\{s, s+\delta, s+2\delta, \ldots,  s+ L -\delta  \}$ with
 $   1 \le s \le \delta $.
 Consider a choice set  $c$  ({\it i.e.},  as explained in section
 \ref{sec:procedure}, a choice of   $N$ roots
 amongst  the $L$ available ones). 
  Suppose  there exist two  packages  ${\mathcal P}_s$ 
  and ${\mathcal P}_t$ such that 
 $$ {\mathcal P}_s  \subset  c  \,\,\,\,\,\,\, {\rm   and   } \,\,\,\,\,\,\, 
     {\mathcal P}_t \cap c = \emptyset    \, . $$
 Consider the  choice set 
 $\hat{c} = ( c\backslash {\mathcal P}_s)\cup {\mathcal P}_t  $
 obtained  by replacing in the  choice set  $c$ the roots  composing
   $ {\mathcal P}_s $  with  the roots   $ {\mathcal P}_t$.
 It can be proved (Golinelli and Mallick 2005b)  that this   choice set
$\hat{c}$ 
 (obtained  from $c$ by exchanging $ {\mathcal P}_s $ and $ {\mathcal P}_t$) 
   corresponds  to the same self-consistent equation   and    to  the
   { \it same eigenvalue} $E$  as $c$.   This property therefore defines
 {\it Equivalence Classes}  amongst the
$ \left(\begin{array}{c} L \\ N\end{array} \right) $ 
 possible  choice sets~:  two  choice sets are equivalent
 if they are obtained from each other  by `Package-swapping'.

\begin{figure}[th]
  \includegraphics[height=7.0cm]{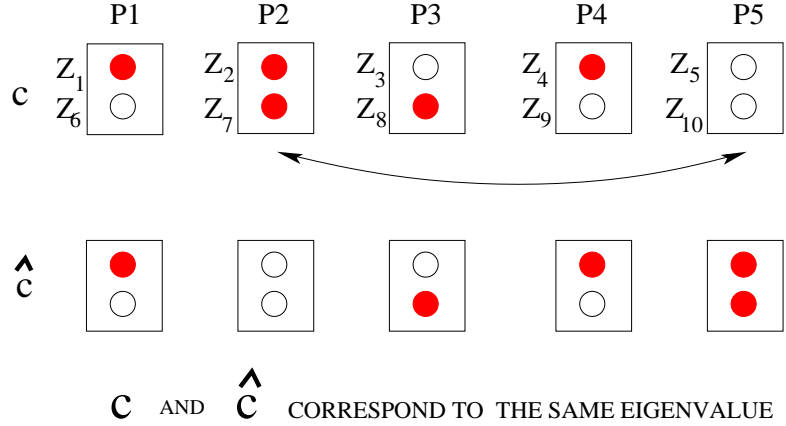} 
 \caption{Example of two equivalent choice functions of the  Bethe roots for 
 the TASEP with  $L=10$  and  $N=5$.}
  \label{fig:packages}
\end{figure}

 Figure \ref{fig:packages} illustrates the case of a system 
 with   $L=10$ sites and  $N=5$  particles, where 
  5 Bethe roots must be chosen  amongst the 10 solutions
 $(Z_1, \ldots Z_{10}) $  lying on a Cassini oval
  of a polynomial of degree 10. Here
 $\delta = \gcd(L,N) = 5$, and therefore the roots can be grouped
 into five different packages~:  
$$ {\mathcal P}_1 = \{Z_1, Z_6\} \, , \,\, 
     {\mathcal P}_2 = \{Z_2, Z_7\} \, , \,\, 
    {\mathcal P}_3 = \{Z_3, Z_8\} \, , \,\, 
     {\mathcal P}_4 = \{Z_4, Z_9\} \, , \,\, 
     {\mathcal P}_5 = \{Z_5, Z_{10}\} \, . $$
 These five packages correspond  to the five  rectangular boxes in figure 
  \ref{fig:packages};  each package contains two roots and the selected
 root is shown as black disk. In this figure, we give the example of 
  a  choice function $c$  that selects  the roots
 $Z_1,Z_2, Z_7,  Z_8, Z_4$. Then, the   choice function $\hat{c}$ that 
  selects   the roots  $Z_1,  Z_8, Z_4, Z_5, Z_{10}$, is such that 
  $\hat{c} = ( c\backslash {\mathcal P}_2)\cup {\mathcal P}_5  $,
 {\it i.e.},   the two choice sets  $c$ and   $\hat{c}$ are 
 obtained from each other by exchanging  the packages 
 ${\mathcal P}_2$ and  ${\mathcal P}_5$. They are thus equivalent
 in the sense  defined above and correspond to the same eigenvalue
 $E = 1/2(-5 + Z_1 + Z_8 + Z_4).$

 The calculation of the degeneracies tables is therefore reduced
 to a pure   problem in combinatorics~: given a choice function $c$, 
 one has to  count   the number of choice functions that are equivalent to $c$,
 and the total number of equivalence classes. Indeed, if we 
 suppose that there is a `one to one'  bijection  between
  choice  sets  and solutions of the Bethe Equations (recall
 that  the  number $\mathcal{ N } $ of  possible 
  choice sets   is the same as the
 dimension of the matrix $M$), then the following  correspondences hold~:
\begin{itemize}
\item[$\bullet$]
 { equivalence  {\bf classes}  by  `package swapping'
     $\leftrightarrow$    {\bf multiplets}  in spectrum }
\item[$\bullet$]
{ {\bf cardinality} }  of a class  $\leftrightarrow$  
{  { \bf order}} of  the multiplet 
\item[$\bullet$]
{{ number}} of  classes of  cardinality $d$ $\leftrightarrow$  
 {{ number}}  of  multiplets  of  degeneracy   $d$.
\end{itemize}

 In the half-filling case, the numbers 
  in  Table  \ref{tab:1/2} are given by  simple formulae
 (Golinelli and Mallick 2004b).  The 
 degeneracy order  $d$ and the number $m(d)$ of multiplets of
 a given degeneracy $d$ can be expressed as a  function
 of a single integer   $r$ 
\begin{equation}
  d_r  = \left(\begin{array}{c} 2r\\r \end{array} \right) , \,\,\, 
   m(d_r)  =   \left(\begin{array}{c} N\\2r \end{array} \right)
   2^{N-2r}     \, ,
 \end{equation} 
 where   $r$ takes all 
 integral values in  the range   $ 0 \le r \le \frac{N}{2} \, .$
 In  arbitrary filling,  expressions for $m$
 and $d$  have  also been  found
 but they depend on $(L/\delta - 1)$  different  parameters  
 (Golinelli and Mallick 2005b).

 \section{Calculation  of large deviation functions}
 \label{sec:LDF}

      The Bethe Ansatz has also led to an exact calculation
 of all the cumulants  of the total current 
 in the TASEP on a ring of size $L$
 with $N$ particles (Derrida and  Lebowitz 1998, Derrida and Appert 1999).
 As explained below, the
  generating function of these cumulants can be expressed as 
 the ground state of a  suitable  deformation of
 the  Markov  matrix. From this  generating function, the 
  large deviation function of the time averaged current
  is obtained exactly. 

 \subsection{A generalized master equation}

  We  call $Y_t$ the total distance
 covered by all the particles between time 0 and time $t$
 and  define $P_t(\mathcal{C}, Y)$ the joint probability
 of being at time $t$ in the configuration $\mathcal{C}$ and
 having $Y_t = Y$. A master equation,
 analogous to equation~(\ref{eq:Markov}), 
 can  be written for  $P_t(\mathcal{C}, Y)$ as follows~:
\begin{equation}
    \frac{d}{dt} P_t(\mathcal{C}, Y)  = \sum_{\mathcal{C}'} \Big(
   M_0(\mathcal{C},\mathcal{C}') P_t(\mathcal{C}', Y)  
+   M_1(\mathcal{C},\mathcal{C}')  P_t(\mathcal{C}', Y -1)  
+   M_{-1} (\mathcal{C},\mathcal{C}')  P_t(\mathcal{C}', Y +1)  
        \Big)   \, . 
 \label{eq:Markov2}
\end{equation}
 In terms of   the generating function  $F_t(\mathcal{C})$ defined as
 \begin{equation}
  F_t(\mathcal{C}) =  \sum_{ Y =0}^\infty 
  {\rm e}^{\gamma Y} P_t(\mathcal{C}, Y) \, ,
 \label{eq:defF}
\end{equation}
 the  master equation~(\ref{eq:Markov2}) takes the simpler
 form~:
\begin{equation}
    \frac{d}{dt} F_t(\mathcal{C})  = \sum_{\mathcal{C}'} \Big(
   M_0(\mathcal{C},\mathcal{C}') 
+ {\rm e}^\gamma  M_1(\mathcal{C},\mathcal{C}')  
+   {\rm e}^{-\gamma}  M_{-1} (\mathcal{C},\mathcal{C}')   
        \Big)  F_t(\mathcal{C}') =  \sum_{\mathcal{C}'} 
  M(\gamma)(\mathcal{C},\mathcal{C}')   F_t(\mathcal{C}')
   \, . 
 \label{eq:Markov3}
\end{equation}
This equation is   similar to the original 
 Markov  equation~(\ref{eq:Markov}) for the probability
 distribution  $P_t(\mathcal{C})$ but where the original
 Markov matrix $M$ is deformed into $ M(\gamma)$  which is given by
 \begin{equation}
  M(\gamma) =   M_0 +  {\rm e}^\gamma  M_1 +  {\rm e}^{-\gamma}  M_{-1} \, . 
 \label{eq:defMgamma}
\end{equation}
 We emphasize  that  $M(\gamma)$,
  that   governs the evolution of  $F_t(\mathcal{C})$, 
  is not a Markov matrix for $\gamma \neq 0$
 (the sum of the elements in  a  given column does not vanish).
 In the long time limit, $ t \to \infty$, 
  the behaviour of   $F_t(\mathcal{C})$
 is dominated by the largest eigenvalue $\lambda(\gamma)$ 
 of the matrix  $M(\gamma)$~:
  \begin{equation}
   F_t(\mathcal{C}) \to  {\rm e}^{\lambda(\gamma) t }
  \langle \, \mathcal{C} \,  | \,  \lambda(\gamma) \,  \rangle \,  , 
 \label{eq:limF}
\end{equation}
 where the ket $ |   \lambda(\gamma)   \rangle $ is the eigenvector
 corresponding to the  largest eigenvalue. 
 Therefore, when  $ t \to \infty$, we obtain  
 \begin{equation}
    \langle  \,  {\rm e}^{\gamma Y_t}\,  \rangle = 
  \sum_{\mathcal{C}}   F_t(\mathcal{C})  \sim  {\rm e}^{\lambda(\gamma)t} 
   \, .
 \label{eq:limF2}
\end{equation}
 More precisely, we have
  \begin{equation}
   \lim_{t \to \infty} \frac{1}{t}
 \log   \langle  \,  {\rm e}^{\gamma Y_t}\,  \rangle =  \lambda(\gamma) 
   \, .
 \label{eq:limF3}
\end{equation}
The function $\lambda(\gamma)$  contains the complete
 information about the  cumulants  of the total current $Y_t$
 in the long time limit.  
  The  large deviation function $G(j)$ for the total current $j$
is  defined as
\begin{equation}
 G(j) =  \lim_{t \to \infty} \frac{1}{t}
 \log \Big[ {\rm Prob}
  \left(      \frac{Y_t}{t} = j \right)   \Big] \, .
 \label{eq:defLDG}
  \end{equation}
 The relation between  $\lambda(\gamma)$ and   $G(j)$ is derived as follows
  \begin{equation} 
 \langle  \,  {\rm e}^{\gamma Y_t}\,  \rangle \sim  {\rm e}^{\lambda(\gamma)t}
 \sim \int  {\rm e}^{ t(G(j) + \gamma j )}  {\rm d}j \, .
\label{eq:LegT}
  \end{equation}
 By saddle-point approximation,   we deduce  that 
 $\lambda(\gamma)$ is the Legendre transform of  $G(j)$
  \begin{equation} 
    \lambda(\gamma) = \max_{j} \Big(  G(j) + \gamma j    \Big) \, .
\label{eq:LegT2}
  \end{equation}
 This relation allows one to obtain the large deviation function $G(j)$
 in the following parametric form, once $\lambda(\gamma)$ is known~:
  \begin{equation} 
   j = \frac{ {\rm d}\lambda}{ {\rm d}\gamma} \,\,\,\, \hbox{ and }
  \,\,\,\ G(j) =   \lambda(\gamma) - \gamma  
  \frac{ {\rm d}\lambda}{ {\rm d}\gamma} \, .
\label{eq:invLeg}
  \end{equation}

 The largest  eigenvalue  $\lambda(\gamma)$  of the 
  deformed matrix   $M(\gamma)$ is  calculated by Bethe
 Ansatz. The   Bethe equations  now read 
\begin{equation}
   z_l^L  = (-1)^{N-1} \prod_{i =1}^N
   \frac{p {\rm e}^{\gamma}  z_l z_i - z_l + q{\rm e}^{-\gamma}  }
  {p{\rm e}^{\gamma}  z_l z_i -z_i +q{\rm e}^{-\gamma} } \,\,\,
  {\rm  for }\,\,\,\,  l =1 \ldots N\, , 
\label{eq:Bethelambda}
\end{equation}
and the corresponding eigenvalue of $M(\gamma)$ is given by
\begin{equation}
 E(\gamma; z_1, z_2 \ldots  z_N ) = p {\rm e}^{\gamma}  \sum_{i =1}^N z_i + 
        q {\rm e}^{-\gamma}  \sum_{i =1}^N \frac{1}{ z_i} - N \, .
\label{eq:vplambda}
\end{equation}

 \subsection{The Gallavotti-Cohen symmetry}

 We remark that the equations~(\ref{eq:Bethelambda}) and
 (\ref{eq:vplambda}) are invariant under the transformation
 \begin{eqnarray}
                 z &\rightarrow& \frac{1}{z}  \nonumber  \\
                 \gamma &\rightarrow& \gamma_0- \gamma 
 \,\,\,\, \hbox{ with } \,\,\,\,   \gamma_0 = \log\frac{q}{p}    \, .
  \end{eqnarray}
 This symmetry implies that the spectrum of  $M(\gamma)$
 and that of  $M(\gamma_0 -\gamma)$  are identical.
 This functional identity  is in particular satisfied by the largest eigenvalue
 of $M$ and we have
\begin{equation}
 \lambda(\gamma) =  \lambda(\gamma_0 - \gamma) \, .
 \label{eq:GC}
  \end{equation}
 Using equation~(\ref{eq:LegT2}), we deduce that this  identity  implies
 the following  symmetry  for the large deviation function 
\begin{equation}
 G(j) =   G(-j) - \gamma_0 j  \, .
 \label{eq:GC2}
  \end{equation}
This relation is a special case of the general {\it Fluctuation
 Theorem} valid for systems far from equilibrium (Evans, Cohen
 and Morriss 1993; Evans and Searles 1994, Gallavotti and Cohen 1995).
 This relation, which manifests itself  here as a symmetry of the
 Bethe equations, does not depend on the integrability
 of the system and can be derived   from
  equation~(\ref{eq:Markov3})  directly
  (Kurchan 1998, Lebowitz and Spohn 1999).

 \subsection{The large-deviation function for the TASEP}

  In the TASEP case, the Bethe equations~(\ref{eq:Bethelambda})
 are simpler and   can be   solved by using
 the procedure outlined in section \ref{sec:procedure}.
An exact  power  series expansion  in $\gamma$ 
 for   $\lambda(\gamma)$ is   obtained by eliminating the parameter $Y$
 in the following two equalities~:
\begin{eqnarray}
      \lambda(\gamma)  &=&  -N   \sum_{k=1}^\infty 
             \left(\begin{array}{c} kL-1\\kN \end{array} \right)
  \frac{Y^k}{kL-1}    \, ,  \label{eq:lamY}\\
           \gamma  &=&  - \sum_{k=1}^\infty   
  \left(\begin{array}{c} kL\\kN \end{array} \right)  \frac{Y^k}{kL} 
        \label{eq:gamY} \, . 
\end{eqnarray}
   Equation~(\ref{eq:gamY}) is inverted 
 by writing  $Y$ formally as a 
   power  series in $\gamma$ and solving recursively for the coefficients
  up to any desired order;  the  function   $\lambda(\gamma)$ is then 
 obtained by substituting  $Y(\gamma)$ in equation~(\ref{eq:lamY}). 

These expressions allow to calculate the cumulants of $Y_t$, for example~:
\begin{eqnarray}
 \lim_{t \to \infty} \frac{ \langle Y_t \rangle}{t} &=&
 \frac{ {\rm d}  \lambda(\gamma)}{ {\rm d} \gamma} 
 \Big|_{\gamma =0} =  \frac{ N(L-N)}{L-1}\, ,    \\ 
 \,\,\, \lim_{t \to \infty} 
\frac{ \langle Y_t^2 \rangle - \langle Y_t \rangle^2 }{t}  &=&
  \frac{ {\rm d}^2  \lambda(\gamma)}{ {\rm d} \gamma^2} 
 \Big|_{\gamma =0} =
 \frac{N^2\; (2L-3)!\; (N-1)!^2 \;(L-N)!^2 }
{ (L-1)!^2 \; (2N-1)! \; (2L-2N-1)! } \, .
 \end{eqnarray}
 In the  limit of a large  system size, $ L \to \infty$,
 and with a fixed  density $\rho = L/N$, these expressions
  reduce to 
\begin{equation}
 \lim_{t \to \infty} \frac{ \langle Y_t \rangle}{t} \simeq
    L \rho(1-\rho)\, ,     \,\,\, \hbox{  and } \,\, \,\,\, 
 \lim_{t \to \infty} 
\frac{ \langle Y_t^2 \rangle - \langle Y_t \rangle^2 }{t}   \simeq
\frac{\sqrt{\pi}}{2}  L^{3/2} [\rho(1-\rho)]^{3/2}  \, .
 \end{equation}
  Using the Stirling formula,  equations~(\ref{eq:lamY}) and~(\ref{eq:gamY})
 take the scaling form   (Derrida and  Lebowitz 1998) 
\begin{equation}
    \lambda(\gamma) - \gamma L \rho(1-\rho)
  =  \sqrt{ \frac{\rho ( 1 -\rho)} {2\pi L^3}} 
 \Phi\left(\gamma \sqrt{2\pi\rho ( 1 -\rho)L^3}  \right)
  \, ,  \label{eq:scalLeg}
 \end{equation}
 where the expression for the  scaling function $\Phi$  does
 not depend on the parameters ($L$ and $\rho$)  of the model.
 From equation~(\ref{eq:scalLeg}),  it can be  shown  that 
  when  $ |j - L\rho ( 1 -\rho) |  \ll L$,  
 the  large deviation function $G(j)$  can  be written as 
\begin{equation}
 G(j) =     \sqrt{ \frac{\rho ( 1 -\rho)} {\pi L^3}}
 H \Big( \frac{j -  L\rho ( 1 -\rho)}{\rho ( 1 -\rho)}   \Big) 
 \label{eq:scalf}
  \end{equation}
with 
\begin{eqnarray}
   H(y) \simeq  - \frac{ 2 \sqrt{3}}{5 \sqrt{\pi}} y ^{5/2} \,\,\,\,
 &\hbox{ for }&  \,\,\,\,   y \to +\infty \, , \\
     H(y) \simeq -  \frac{ 4 \sqrt{\pi}}{3} |y| ^{3/2} \,\,\,\,
 &\hbox{ for }&  \,\,\,\,   y \to -\infty \, . 
\end{eqnarray}
 This distribution is skew: it decays as  a  power law
 with an exponent $5/2$ for  $y \to +\infty$ 
 and with an exponent $3/2$ for  $y \to -\infty$.  

 \section{Applications to related  models}
 \label{sec:variantes}

   In the previous sections, we have  put   emphasis 
 on  the Bethe Ansatz solution of 
  the totally asymmetric exclusion process on a ring.  In this special case,
 the  Bethe equations have a particularly simple
 form and their analysis can be carried out thoroughly. In this
 last section, we briefly present results that have been obtained
  with  the  Bethe Ansatz for the  partially
  asymmetric exclusion process (PASEP), 
 for the exclusion process   with an impurity,  and for the exclusion process
 with open boundaries. 

 \subsection{The partially asymmetric exclusion process}

       The Bethe equations~(\ref{eq:Bethepq}) for general hopping
 rates  $p$ and $q$ can not be reduced to an effective  one-variable
 polynomial equation as in the TASEP case and their analysis  becomes
 much more involved. In particular,  combinatorial expansions 
 for the energy such as (\ref{eq:CombE}) valid for finite values
 of $L$ and $N$ are not known.  However,  a very thorough study 
 has been carried out, in the limit $L \to \infty$, by Doochul Kim
 and his collaborators in a series of papers
  (Noh and Kim 1994, Kim 1995, Kim 1997, 
 Lee and Kim 1999; see also related works on the non-Hermitian XXZ chain by
  Albertini et al. 1996, 1997).
  They developed a perturbative scheme that
 enabled  them to calculate the finite size corrections
 of the gap and the  low lying excitations of the asymmetric
 XXZ chain.  They derived  the crossover scaling functions of the energy
 gaps from the symmetric case $p = q = 1/2$ to the asymmetric 
 case. In the continuous limit, these  functions   describe
 a crossover from the Kardar-Parisi-Zhang universality class
 to the Edwards-Wilkinson class.

  The same technique leads  to the
 calculation of the large deviation function of the PASEP
   in the scaling limit (Lee and Kim 1999).  The large
 deviation function  takes the form
\begin{equation}
 G(j) =   (p-q)  \sqrt{ \frac{\rho ( 1 -\rho)} {\pi N^3}}
 H \Big( \frac{j -  L (p-q)\rho ( 1 -\rho)}{ (p-q)\rho ( 1 -\rho)}   \Big) 
 \label{eq:scalfpq}
  \end{equation}
 where the function $H$ is the same as that of 
  equation~(\ref{eq:scalf}). The only modification that
 occurs for  the PASEP  is the  rescaling factor $p-q$. 
 Non-trivial differences appear in the subleading terms, but these
 corrections are likely to be model-specific and non-universal. 
  Using  this   expression of the
 large deviation function, it  is possible to verify explicitly 
  that the Gallavotti-Cohen relation~(\ref{eq:GC2})
 is satisfied (Lebowitz and Spohn 1999).  However, an  exact
   power series   expansion  of 
 $\lambda(\gamma)$    in terms
 of the deformation parameter $\gamma$ is not known.   Such
 an  expansion would lead to exact formulae for 
 the cumulants of the current valid  for any values of $L$ and $N$.
 We believe, nevertheless, that  expressions
  analogous to equations~(\ref{eq:lamY} and \ref{eq:gamY})
 should exist for the PASEP because exact combinatorial
 formulae for the mean value  of the current and its variance  have  
 been calculated by using the Matrix Product
 method   (Derrida and Mallick  1997).

 \subsection{ASEP  with an impurity}

 The presence of a defective particle  (say, an impurity)
 in  the ASEP on a ring can generate a shock dynamically
 in the stationary state (Mallick, 1996). A traffic-flow
 picture  nicely illustrates this fact~: if particles
 represent  cars and the impurity a truck (which moves
 at  a slower speed and is difficult to overtake), then
 the shock corresponds to a traffic jam. The phase transition
 to a shock can also be interpreted as  a real space version
 of Bose-Einstein condensation (Evans 1996).

\begin{figure}[th]
  \includegraphics[height=5.0cm]{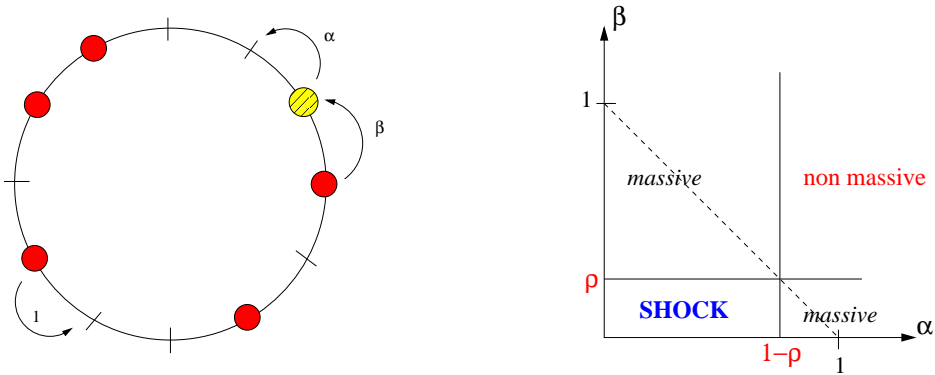}
   \caption{\em  The Asymmetric exclusion process on a periodic
 ring with an impurity.}
  \label{fig:ASEPimpurity}
\end{figure}

 The model with a single defect particle is drawn 
 in Figure \ref{fig:ASEPimpurity}. The system  contains $N$
 normal particles (denoted by 1) and one impurity (denoted by 2).
 Each site is either occupied by a particle or by the impurity,
 or it is empty.  The stochastic dynamical rules that
 govern the evolution of the system during the infinitesimal
 time step $dt$ are 
 \begin{eqnarray}
      10 &\rightarrow& 01 \,\,\,\,\, \hbox{ with rate } \,\,\,\,\, 1
 \nonumber \\
     20 &\rightarrow& 02 \,\,\,\,\, \hbox{ with rate } \,\,\,\,\, \alpha \\
     12 &\rightarrow& 21 \,\,\,\,\, \hbox{ with rate } \,\,\,\,\,  \beta \, .
 \nonumber
 \end{eqnarray}
 All other transitions are forbidden.  The phase diagram of this
 model can be determined exactly by the Matrix product method 
 and  consists of  four main phases (see  Figure \ref{fig:ASEPimpurity}).
 In the two  massive phases, the perturbation due to the  impurity 
  is of finite range and all correlations decay
 exponentially. In the non-massive phase, the impurity has a
 long range effect~: correlations  decay algebraically. In the shock
 phase, the presence of the unique impurity induces a phase
 separation between  a low density region and a  high density  region.

 By using the Bethe Ansatz, Derrida and Evans (1999) have
 calculated  the large deviation function of the
 total displacement of the defect particle. Their result  leads to analytical
 formulae for 
 the velocity and the diffusion constant of the impurity in the various
 phases. In particular, in the shock phase, the location  of
 the defect  can be identified with the position  of the shock and
 the  large deviation function  provides  complete
 information on  its statistical
 behaviour.

 A more direct way to produce a shock  in a  system 
  is to introduce a slow bond
  that particles cross with a rate $r < 1$ (Janowski
 and Lebowitz 1992, 1994). For $r$ less than a critical value
 that depends on the mean  density of particles, 
 the system exhibits a phase separation. This model, in spite
 of many efforts,  has not been solved exactly and is probably not 
 an integrable system (it does not satisfy any  known
 criterion  for integrability). However,  G. Sch\"utz (1993)
 has solved, using  Bethe Ansatz,  a variant
 of this model in which  particles move deterministically 
 on a ring of length $L$ (with $L$ even) with a single defect
 across which the particles jump with probability  $r < 1$. 
 In this dynamics, the time evolution consists
 of two intermediate  time steps in which  even sites and odd sites 
 are updated simultaneously. 

 \subsection{The open boundary model}

\begin{figure}[th]
  \includegraphics[height=2.75cm]{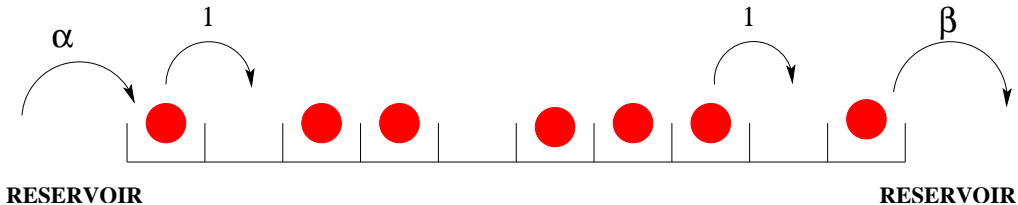}
   \caption{The Totally  Asymmetric exclusion process on an open  
chain}
  \label{fig:openASEP}
\end{figure}

   The asymmetric exclusion process with open boundaries
   can be viewed as a model for a driven lattice gas in contact
 with two reservoirs. In the bulk, the dynamics of the ASEP is the 
 same as that  shown in Figure 1; at the boundaries particles
 can enter or leave the system with various input/output rates.
 The totally asymmetric  version of this model is illustrated 
 in Figure~\ref{fig:openASEP}~: particles in the bulk  jump to
 to right with rate 1; a particle can be  added 
 with rate $\alpha$  if the site 1 is empty;
  a particle can  be removed    with rate $\beta$ 
 if the site L is occupied.

 The open  ASEP  undergoes boundary induced
 phase  transitions (Krug 1991). The density profile and the current 
 can be calculated exactly by using
 a Matrix product representation for the steady
 state (Derrida et al. 1993; see also
 Sch\"utz and Domany 1993 for an alternative derivation). In the large
 system size limit, the expressions for the current and the density
 profile become non-analytic  for certain values
 of  the input/output rates. This  leads to   a phase diagram
 with three main regions~: a low density phase (when  typically 
 the input rates are small and the output rates
 are large),     a high density  phase (when  
   the input rates are  large and the output rates small), 
 and a maximal current phase (in which the bulk density   is
 1/2 regardless of the boundary rates). 
     
 Although  the ASEP with open boundaries was known to be an integrable
 system (Essler and Rittenberg 1996),  the explicit  diagonalisation
 of the Markov matrix by   Bethe Ansatz could not be carried
 out because of the  boundary  terms that violate the 
 conservation  of the total number of particles. Only   numerical
 and phenomenological  studies  of the spectrum were available
 (Bilstein and  Wehefritz  1997,  Dud\-zi\'nski and Sch\"utz  2000).
  In a  remarkable recent  paper, de Gier and Essler (2005)
 were able to derive the Bethe Ansatz equations describing the
 complete spectrum of the Markov matrix of the 
 partially asymmetric exclusion process with open boundaries and 
 general input/output rates. Their work  uses  an  
  exact solution of the open  XXZ chain with non-diagonal boundary terms
(Nepomechie 2003 and 2004, Cao et al. 2003).
 De Gier and Essler have 
  calculated   analytically the spectral gap  of the TASEP
 with open boundaries and have  discussed  the various regimes throughout
 the phase diagram; in particular they have discovered    a fine  structure
 of   the phase diagram  that  could not be found from the stationary
 state alone. Their work 
 opens the way to  many further investigations,
 such as the description of the excited states and the calculation
 of the large deviation function of the current in the open system. 

 \section{Conclusion}

    The Bethe Ansatz was initially used to diagonalize 
 Quantum Hamiltonians  (such as the Heisenberg Spin Chain,
 the one-dimensional Bose gas with $\delta$-interactions)
 and to  calculate real spectra of Hermitian Operators. In the 70's
 and the beginning of the 80's, 
 the exact solutions of the six and eight vertex models and the
 development of the quantum inverse scattering method
 (Baxter 1982, Faddeev 1984) showed that many systems of interest 
 in equilibrium statistical mechanics belonged to the class of integrable
 models.  The ASEP, which plays the role of a paradigm in non-equilibrium
 statistical mechanics,  thanks to its  rich and complex  phenomenology,  has  
  been shown to be  integrable.  This integrability property has allowed
 to derive many exact solutions  for the ASEP and for some related models
 that we have reviewed here.  These analytical  results  add to our 
  knowledge of systems far from equilibrium  and allow us
 to compare them with  systems at equilibrium  on a quantitative level. 

 The use of integrability techniques
 in non-equilibrium statistical mechanics is an  active 
  field of current research. In particular, we believe that 
 the following  problems should lead to interesting results in  near future~:
  the use of  the nested  Bethe Ansatz for ASEP with multiple
 species of particles,  the  investigation  of   ASEP   with open boundaries,
   the study of  exclusion  processes with disorder, and
  the application of integrability techniques to the  zero-range process
  (Povolotsky 2004,   Povolotsky and Mendes 2006)

\subsection*{Acknowledgments}
   We wish to thank the guest editors of this special issue,
 Patrick  Dorey,   Gerald  Dunne and  Joshua  Feinberg for
 inviting us to write this contribution. We are particularly
 thankful to  Patrick  Dorey for his encouragements and his patience. 
 We thank  B. Derrida,   M. Gaudin,  V. Hakim, T. Jolic{\oe}ur,  S. Majumdar, 
 S. Nechaev,  V. Pasquier, G. Sch\"utz, H. Spohn,
  R. Stinchcombe and V. Rittenberg
 for  useful discussions.  We are particularly greatful 
  to S. Mallick for a critical reading of the manuscript. 
 We thank the anonymous referee for many  very helpful  comments.

\section*{References}
\begin{itemize}

\item 
 G.~Albertini, S.~R.~Dahmen, B.~Wehefritz, 1996, 
{\em Phase diagram of the non-Hermitian asymmetric XXZ  spin chain}, 
J. Phys. A: Math. Gen. {\bf 29}, L369.

\item 
 G.~Albertini, S.~R.~Dahmen, B.~Wehefritz, 1997,
{\em The free energy singularity of the asymmetric 6-vertex model
 and the excitations of the asymmetric XXZ  chain},
 Nucl. Phys. B {\bf 493}  541.

\item 
F.~C.~Alcaraz, M.~Droz, M.~Henkel, V. Rittenberg,  1994,
{\em Reaction-diffusion processes, critical dynamics,  and quantum
 chains},
 Ann. Phys.  {\bf 230}, 250.

\item 
F.~C.~Alcaraz,  M.~J.~Lazo, 2004,
{\em The Bethe Ansatz as a matrix product Ansatz}, 
J. Phys. A: Math. Gen. {\bf 37}, L1-L7

\item 
 R.~J.~Baxter, 1982  
 {\em Exactly solvable models in Statistical Mechanics}
(Academic Press, San Diego).

\item 
 H. Bethe, 1931,
 {\em Zur Theory der Metalle. I Eigenwerte und Eigenfunctionen
 Atomkette}  Zur Physik  {\bf 71}, 205. 

\item 
U. Bilstein, B. Wehefritz, 1997, 
 {\em  Spectra of non-Hermitian quantum spin chains describing
 boundary induced phase transitions}, 
J. Phys. A: Math. Gen.  {\bf 30}, 4925.

\item 
 R. Bundschuh, 2002, 
{\em Asymmetric exclusion process and extremal statistics of
 random sequences},
Phys. Rev. E {\bf 65}, 031911.

\item 
  J.~Cao, H.-Q.~Lin, K.-J.~Shi, Y.~Wang, 2003,
 {\em Exact solution of XXZ spin chain with unparallel boundary fields},
 Nucl. Phys. B.  {\bf 663}, 487.

\item 
 B. Derrida, 1998, 
{\em An exactly soluble non-equilibrium system: the asymmetric simple
 exclusion process},  
 Phys. Rep.  {\bf 301}, 65.

\item 
  B. Derrida, C. Appert, 1999,
{\em Universal large-deviation function of the Kardar-Parisi-Zhang
 equation in one dimension},  
 J. Stat. Phys. {\bf 94}, 1. 

\item 
 B. Derrida, M. R. Evans, 1999,
 {\em Bethe Ansatz solution for  a defect particle in 
the asymmetric exclusion process}, 
 J. Phys. A: Math. Gen. {\bf 32}, 4833.

\item 
 B. Derrida, M.~R.~Evans, V. Hakim, V. Pasquier, 1993,
 {\em Exact solution of a 1D  asymmetric exclusion
model  using a matrix formulation}, 
J. Phys. A: Math. Gen. {\bf 26}, 1493.

\item 
 B. Derrida, J. L. Lebowitz, 1998,
{\em Exact  large deviation function in the asymmetric exclusion process},  
 Phys. Rev. Lett.  {\bf 80}, 209.

\item 
 B. Derrida,  J.~L. Lebowitz,  E.~R. Speer, 2003, {\em Exact large
    deviation functional of a stationary open driven diffusive system:
    the asymmetric exclusion process}, J. Stat. Phys. {\bf 110}, 775.

\item   
 B. Derrida, K. Mallick, 1997, {\em Exact diffusion constant 
 for the  one-dimensional  partially asymmetric exclusion process}, 
        J. Phys. A: Math. Gen.   {\bf  30}, 1031.

\item 
  D. Dhar, 1987,
{\em    An exactly solved model for interfacial growth},
  Phase Transitions {\bf 9}, 51.

\item 
M. Dudzi\'nski, G.~M.~Sch\"utz, 2000,
  {\em Relaxation spectrum of the asymmetric exclusion
  process with open boundaries},
 J. Phys. A: Math. Gen. {\bf 33}, 8351.

\item 
 F.~H.~Essler,  V. Rittenberg,  1996
 {\em  Representations of the quadratic algebra and partially
 asymmetric diffusion with open boundaries}, 
 J. Phys. A: Math. Gen. {\bf 29}, 3375.

\item  
  M. R. Evans, 1996,  {\em Bose-Einstein condensation in
 disordered exclusion models and relation to traffic flow}, 
 Europhys. Lett. {\bf 36}, 13.

\item 
 D.~J.~Evans, E.~G.~D.~Cohen, G.~P.~Morriss, 1993  
 {\em Probability of Second Law Violations
 in Shearing Steady states},   Phys. Rev. Lett.  {\bf 71}, 2401.

\item 
 D.~J.~Evans, D.~J.~Searles,  1994  
 {\em  Equilibrium microstates which generate
 second law violating steady states},   Phys. Rev. E  {\bf 50}, 1645.

\item 
 L.~D.~Faddeev, 1984, 
 {\em Integrable models in 1+1 dimensional quantum field theory},
 in  Les Houches Lectures  1982 (Elsevier).

\item
 G.~Gallavotti,  E.~G.~D.~Cohen,  1995
 {\em  Dynamical ensembles in non-equilibrium statistical mechanics}, 
   Phys. Rev. Lett.  {\bf 74}, 2694; 
{\em  Dynamical ensembles in stationary states},
 J. Stat. Phys  {\bf 80}  931.

 \item 
  J.~de~Gier, F.~H.~L. Essler, 2005,
  {\em  Bethe Ansatz solution of the Asymmetric Exclusion Process
 with Open Boundaries}, Phys. Rev. Lett.  {\bf 95}, 240601.

\item O. Golinelli, K. Mallick, 2004a,
 {\em Bethe Ansatz calculation of the spectral gap of the
 asymmetric  exclusion process},  
 J. Phys. A: Math. Gen.  {\bf 37}, 3321.

\item  O.  Golinelli,  K.  Mallick, 2004b,
 {\em  Hidden symmetries in the asymmetric
  exclusion process},    J. Stat. Mech.: Theor. Exp.
 P12001468/2004/12/P12001.

\item O.  Golinelli,  K.  Mallick, 2005a,
 {\em  Spectral gap of the totally
 asymmetric exclusion process at arbitrary filling},  J. Phys. A: Math. Gen.
   {\bf 38}   1419.

\item O.  Golinelli,  K.  Mallick, 2005b,
{\em Spectral Degeneracies in the  Totally Asymmetric Exclusion Process}, 
  J. Stat. Phys  {\bf 120}  779.

\item O.  Golinelli,  K.  Mallick, 2006,
{\em  Derivation of a matrix product representation for the
 asymmetric exclusion process from Algebraic Bethe Ansatz},
  J. Phys. A: Math. Gen.  {\bf 39}   10647.

\item  L.-H. Gwa, H. Spohn, 1992,
  {\em Bethe solution for the dynamical-scaling exponent of the noisy
  Burgers equation},
  Phys. Rev. A {\bf 46}, 844.

\item  T. Halpin-Healy, Y.-C.~Zhang, 1995, 
{\em Kinetic roughening phenomena, stochastic growth, directed polymers and
 all that}, Phys. Rep.  {\bf 254}, 215.

\item  S.~A.~Janowski,   J.~L.~Lebowitz, 1992 
 {\em  Finite size effects and Shock fluctuations in the asymmetric
 exclusion process},   Phys. Rev. A  {\bf 45}, 618.

\item  S.~A.~Janowski,   J.~L.~Lebowitz, 1994 
 {\em  Exact results for the  asymmetric
 exclusion process with a blocage},    J. Stat. Phys.  {\bf 77}, 35.

\item  D.~Kandel, E.~Domany,  B.~Nienhuis, 1990,
   {\em A six-vertex model as a diffusion problem: derivation of
 correlation functions}, 
 J. Phys. A: Math. Gen.  {\bf 23}, L755.

 \item S.~Katz, J.~L.~Lebowitz, H.~Spohn, 1984,
  {\em  Nonequilibrium steady states of stochastic lattice gas models
 of fast ionic conductors},
 J. Stat. Phys.  {\bf 34}, 497.

 \item    D. Kim, 1995,
 {\em Bethe Ansatz solution  for crossover scaling functions
 of the asymmetric XXZ chain  and the Kardar-Parisi-Zhang-type
 growth model},
 Phys. Rev. E {\bf 52}, 3512.

\item   D. Kim, 1997,  
 {\em Asymmetric  XXZ chain at the antiferromagnetic transition:
 spectra and partition functions},  
 J. Phys. A: Math. Gen.  {\bf 30}, 3817.

 \item   S.~Klumpp, R.~Lipowsky, 2003,
 {\em Traffic of molecular motors through tube-like compartments},
  J. Stat. Phys.  {\bf 113}, 233.

 \item   J.~Krug, 1991,
  {\em Boundary-Induced Phase Transitions in Driven Diffusive
 Systems},  
 Phys. Rev. Lett.  {\bf 67}, 1882.

\item 
J. Krug, 1997,
 {\em Origins of scale invariance in growth processes}, 
 Adv.   Phys. {\bf 46}, 139.

\item J.~Kurchan, 1998, 
{\em  Fluctuation theorem for stochastic dynamics},
 J. Phys. A: Math. Gen.  {\bf 31}, 3719.

\item J. L. Lebowitz, H.~Spohn, 1999 
 {\em A Gallavoti-Cohen type symmetry in the large
 deviation functional for stochastic dynamics},  J. Stat. Phys.  {\bf 95}, 333.
 
 \item   D.~S.~Lee, D. Kim, 1999,
 {\em Large deviation function of the partially asymmetric
 exclusion process}, 
 Phys. Rev. E {\bf 59}, 6476.

\item  D. G. Levitt, 1973,
 {\em Dynamics of a single-file pore: Non-Fickian behavior}, 
 Phys. Rev. A {\bf 8}, 3050.

\item  T.~M.~Liggett, 1985, 
{\em Interacting Particle Systems}, (Springer-Verlag, New-York).

\item  T.~M.~Liggett, 1999, 
{\em  Stochastic Models of Interacting  Systems:Contact, Voter
 and Exclusion Processes}, (Springer-Verlag, New-York).

\item   C.~T.~MacDonald, J.~H.~Gibbs,  A.~C.~Pipkin, 1968,
   {\em Kinetics of biopymerization on nucleic acid templates},
 Biopolymers {\bf 6}, 1. 

\item   C.~T.~MacDonald, J.~H.~Gibbs, 1969,
  {\em Concerning the kinetics of poly\-peptide synthesis
 on polyribosomes}, 
Biopolymers {\bf 7}, 707.

 \item   K.~Mallick, 1996,
 {\em Shocks in the asymmetric  exclusion model with an impurity},
   J. Phys. A: Math. Gen.  {\bf 29}, 5375.

 \item   R.~I.~Nepomechie, 1999,
 {\em A spin Chain Primer}, Int.J.Mod.Phys.B  {\bf 13}, 2973.

 \item   R.~I.~Nepomechie, 2003,
 {\em Functional Relations and Bethe Ansatz for the XXZ Chain},
 J.  Stat. Phys.  {\bf 111}, 1363.  

 \item   R.~I.~Nepomechie, 2004,
 {\em Bethe Ansatz solution of the open XXZ chain with nondiagonal
 boundary terms},
 J. Phys. A: Math. Gen.  {\bf 37}, 433.

 \item   J.~D.~Noh, D.~Kim, 1994,
 {\em  Interacting domain walls and the five-vertex model},
   Phys. Rev. E {\bf 49}, 1943.

\item   A.~M.~Povolotsky, 2004,
  {\em   Bethe Ansatz solution of zero-range process with
 nonuniform stationary state},  
Phys. Rev. E {\bf 69}, 061109.

\item   A.~M.~Povolotsky, J.~F.~F.~Mendes, 2006,
 {\em Bethe ansatz solution of discrete time stochastic processes
 with fully parallel update}, 
 to appear in  J. Stat. Phys.

\item   V.~B.~Priezzhev, 2003,
 {\em Exact Nonstationary Probabilities in the  Asymmetric Exclusion Process
   on a Ring},
  Phys. Rev. Lett.  {\bf 91}, 050601.

\item   A.~R\'akos,  G.~M.~Sch\"utz, 2005,
 {\em   Bethe Ansatz and current distribution for the TASEP
 with particle-dependent hopping rates},  cond-mat/0506525

\item   P. M. Richards, 1977,
 {\em Theory of one-dimensional hopping conductivity and diffusion}, 
 Phys. Rev. B {\bf 16}, 1393.

 \item   T.~Sasamoto,  M.~Wadati, 1998, 
  {\em  Exact results for one-dimensional totally asymmetric 
 diffusion models}, J. Phys. A: Math. Gen.  {\bf 31}, 6057.

\item   B.~Schmittmann and R.~K.~P. Zia, 1995,
 {\em  Statistical mechanics  of driven diffusive systems}, 
  in {\em Phase Transitions and Critical Phenomena vol 17.}, C. Domb and
 J.~L.~Lebowitz Ed., (San Diego, Academic Press).

\item   M.~Schreckenberg, D.~E.~Wolf (ed.), 1998,  
  {\em  Traffic and granular flow '97}
 (Springer-Verlag, New-York).

\item   G.~M.~Sch\"utz, 1993,
 {\em Generalized  Bethe Ansatz solution of a one-dimen\-sional
  asymmetric exclusion  process on a ring with blockage},
 J. Stat. Phys. {\bf 71},  471.

 \item   G.~M.~Sch\"utz, 1997,
 {\em Exact solution of the master equation of the asymmetric
 exclusion process},
 J. Stat. Phys. {\bf 88},  427.

\item   G.~M.~Sch\"utz, 2001,
{\em Exactly Solvable Models for Many-Body Systems Far from Equilibrium}
 in {\em Phase Transitions and Critical Phenomena vol 19.}, C. Domb and
 J.~L.~Lebowitz Ed., (Academic Press, San Diego).

 \item   E.~R. Speer, 1993,
{\em The two species totally  asymmetric exclusion  process},
 in Micro, Meso and Macroscopic approaches in Physics, M.~Fannes
 C. Maes and  A. Verbeure Ed. NATO Workshop 'On three levels',
 Leuven, July 1993.

\item   F. Spitzer, 1970,
 {\em Interaction of Markov Processes}, Adv. in Math. {\bf 5},  246.

\item  H. Spohn, 1991,
{\em Large scale dynamics of interacting particles},
 (Springer-Verlag, New-York).

\item R.~Stinchcombe \,  G.~M.~Sch\"utz, 1995, 
 {\em Application of Operator algebras to stochastic dynamics
 and the Heisenberg chain},   Phys. Rev. Lett.   {\bf 75}, 140. 

\item 
H.~J.~de~Vega,  1989,
 {\em Yang-Baxter algebras, integrable theories and quantum groups},
  Int. Jour. Mod. Phys. A {\bf 4}, 2371.

\item  B.~Widom, J.~L.~Viovy, A.~D.~Defontaines, 1991,
{\em  Repton model of gel electrophoresis and diffusion},
J. Phys. I France {\bf 1},  1759.

\end{itemize}

\end{document}